# Tailoring defects and nanocrystal transformation for optimal heating power in bimagnetic $Co_yFe_{1-y}O@Co_xFe_{3-x}O_4$ particles

George Antonaropoulos,[a,b] Marianna Vasilakaki,[c] Kalliopi N. Trohidou,[c] Vincenzo Iannotti,[d] Giovanni Ausanio,[d] Milinda Abeykoon,[e] Emil S. Bozin[f] and Alexandros Lappas *[a]

*Institute of Electronic Structure and Laser, Foundation for Research and Technology - Hellas, Vassilika Vouton, 71110 Heraklion, Greece*

*Department of Chemistry, University of Crete, Voutes, 71003 Heraklion, Greece*

*Institute of Nanoscience and Nanotechnology, National Center for Scientific Research Demokritos, 15310 Athens, Greece*

*CNR-SPIN and Department of Physics "E. Pancini", University of Naples Federico II, Piazzale V. Tecchio 80, 80125 Naples, Italy*

*Photon Sciences Division, National Synchrotron Light Source II, Brookhaven National Laboratory, Upton, New York 11973, USA*

*Condensed Matter Physics and Materials Science Division, Brookhaven National Laboratory, Upton, New York 11973, USA*

The effects of cobalt incorporation in spherical heterostructured iron oxide nanocrystals (NCs) of sub-critical size have been explored by colloidal chemistry methods. Synchrotron X-ray total scattering methods suggest that cobalt (Co) substitution in rock salt iron oxide NCs tends to remedy its vacant iron sites, offering a higher degree of resistance to oxidative conversion. Self-passivation still creates a spinel-like shell, but with higher volume fraction of the rock salt Co-containing phase in the core. The higher divalent metal stoichiometry in the rock salt phase, with increasing Co content, results in a population of unoccupied tetrahedral metal sites in the spinel part, likely through oxidative shell creation, involving an ordered defect-clustering mechanism, directly correlated to the core stabilization. To shed light on the effects of Co-substitution and atomic-scale defects (vacant sites), Monte Carlo simulations suggest that designed NCs, with desirable, enhanced magnetic properties (cf. exchange bias and coercivity), are developed with magnetocrystalline anisotropy raised at relatively low content of Co ions in the lattice. Growth of optimally performing candidates combines also a strongly exchange-coupled system, secured through a high volumetric ratio rock salt phase, interfaced by a not so defective spinel shell. In view of these requirements, Specific Absorption Rate (SAR) calculations demonstrate that the sufficiently protected from oxidation rock salt core and preserved over time heterostructure, play a key role in magnetically-mediated heating efficacies, for potential use of such NCs in magnetic hyperthermia applications.

*Corresponding author, lappas@iesl.forth.gr

## Introduction

Magnetic Nanoparticles (MNPs), a sub-class of nanoscale crystals (nanocrystals: NCs)[1] that consist of open-shell transition metal elements, have been the subject of considerable research endeavor in recent years. Their promises stem from their capability to respond to electromagnetic stimuli (e.g. a magnetic field), thus enabling a broad spectrum of applications ranging from catalysis,[2] photonics,[3] and data storage,[4] to water treating,[5] energy storing[6] and printing technologies[7]. Moreover, from the standpoint of view of the present work, MNPs offer highly exploitable capabilities in the rapidly developing field of nanobiotechnology.[8] The latter, amongst others, includes medical diagnostics with contrast agents, for magnetic resonance imaging (MRI) technologies and therapeutic strategies with heat emission carriers, for magnetic hyperthermia treatments, both taking advantage of facile bottom-up pathways for controlling and boosting the nanoparticles' magnetic performance.[9] In this endeavor, biocompatibility and low toxicity are necessary, but understanding how structural and morphological characteristics can be utilized as materials' design parameters, still require insights to improve the MNPs' performance. More specifically, magnetic heating under the influence of alternating (AC) magnetic fields depends on Néel-Brown relaxations, which evolve with magnetic anisotropy constant ($K$) and saturation magnetization ($M_s$),[10] thus offering a pathway for hyperthermia efficiency optimization. Changing the NCs' geometrical parameters, such as size[11] and shape,[12] offer another avenue to tune the associated relaxation times. While bottom-up, colloidal chemistry approaches are exploited with the purpose to imprint the afore-mentioned materials-design characteristics in favor of hyperthermia, top-down fabrication methods offer alternative exploitable morphologies. The derived MNPs, may entail either a multi-layered structure called synthetic antiferromagnet (SAF), or involve a single magnetic layer having a magnetic vortex configuration.[13] Interestingly, such artificial MNPs mimic the superparamagnetic ones made by chemical synthesis, but with advantages when dispersion and actuation in biological media is required. Furthermore, the bottom-up growth of topologically distinct phases, such as in a core@shell-like geometry,[14] involving contrasting magnetic-ordering states between the core and the shell [e.g. antiferromagnetic (AFM), ferromagnetic (FM) and/or ferrimagnetic (FiM)], has proven to be particularly effective in introducing exchange anisotropies. Such systems, take advantage of the technologically exploitable exchange-bias effect,[15,16] with concomitant enhancement of the NCs' hyperthermia response.[12] Last but not least, composition also plays a key role in changing the MNPs' magnetocrystalline anisotropy[17], thus allowing for further regulation of the Néel-Brown relaxation behaviour. In this respect, chemical substitution strategies have been evaluated, in an effort to tune the magnetic response of iron oxide NCs.[18,19]

## Compositional control of NCs

With the purpose to either provide new, or enhance existing properties (cf. mechanical, electrical, optical, biological, magnetic etc.), changing the chemical composition by introducing structurally compatible ions, has become a widely used method for the modification of the crystal lattice of a number of materials. Such an adjustment is being achieved through the incorporation of usually a small amount of atoms or ions, which would not normally be present in the matrix of the parent material crystal lattice. The ion being incorporated, can be either introduced in an interstitial crystallographic site or, most usually in the case of NCs, it can substitute another ion from the matrix.[20] A special case of crystal lattice chemical alteration is doping, where ions of the matrix are replaced with ions of the same[21] or different number of valence electrons,[22] aiming mainly at the modification of their electronic structure physical properties.[23] The effects of such chemical substitutions become particularly noticeable in the nanoscale and great progress has been achieved in recent years, concerned with the control of chemical composition in nanocrystalline materials. Epitaxial growth of doped NCs gave great results in the past, allowing for significant technological development in novel materials, such as quantum dots.[24] On the other hand, fabrication of doped colloidal NCs offers additional advantages, as for example, the opportunity to control the photoluminescence of colloidal semiconductor NCs, raising at the same time the technological challenges.[25] Thus, over the last 15 years, research efforts have focused on ionic substitution in nanocrystal by means of colloidal chemistry principles. In this effort, two main pathways can be identified, substitution through ion diffusion or by mixing the appropriate precursors. In the first pathway, pre-formed colloidal NCs are being further subjected to an additional synthetic step, which includes a colloidal solution, containing the dopant/additive ions. For bulk materials, this demands elevated temperatures, as the procedure is governed by diffusion effects. The same procedure, for nanocrystalline derivatives on the other hand, can take place at moderate temperatures, typically lower than the temperatures routinely used in high-T colloidal syntheses. In the nanoscale, given the increased surface to volume ratio, the predominant effect is the absorption of the added ion at the highly active surface of the nanocrystal. Such procedures are known as ion-exchange methods, as the "additive" replaces an anion or cation accordingly, in the host crystal lattice. Their success is determined by the accurate control of four distinct processes: surface absorption of the new ion, incorporation in the crystal lattice of the parent phase, lattice diffusion, and lattice ejection. [26] As that late years have seen a surge of interest in using this method to form otherwise hard to obtain nanocrystal morphologies, with fine-tuned properties.[27] The second pathway, requires mixing all the ingredients necessary for the fabrication of the final product at the initial stage of the chemical reaction. Precursors which contain the necessary ions for the synthesis (e.g. metal-organic compounds), surfactant molecules and an appropriate solvent, are typical ingredients for a colloidal nanoparticle synthesis. This assumes that both the ions of the host material and the substitutes are chemically compatible and capable to form precursors, which react under the specified experimental conditions in a similar and predictable way. Choosing one over the other approach for the growth of core@shell iron oxide NCs, are further discussed in the subsequent experimental section.

## Advantages of chemical substitution in MNPs

The challenges for optimal nanomaterials' composition, occasionally come along with questions of how efficiently MNPs generate heat under AC magnetic field excitations. As magnetic hyperthermia is considered an innovative approach in synergistic therapeutic avenues for cancer treatment,[28,29] materials' design pathways to improve the hyperthermic properties are needed. One common strategy is to try to increase the magnetic anisotropy. At the same time, size is a critical factor for MNPs, especially for biomedical applications, where minimal amounts of relatively small yet magnetically active particles should be used to avoid possible cytotoxicity. Thus, designed MNPs ideally would rest in the superparamagnetic regime (i.e. at critical sizes, d< 20-30 nm),[9,30] without compromising much of their magnetic power. Many promising strategies have been explored in this direction, including, modification of morphological features (size-shape tuned; d~ 20-40 nm),[31] growth of shells (for interfacial interactions; d~ 15 nm)[32,12] and alteration of chemical composition (e.g. Zn-substituted ferrites; d~ 15 nm).[33] The efficiency of the first two strategies, together with favourable nano-structural effects due to emerging Fe-site vacancies, mediating the composition, have been recently discussed in the $Fe_xO@Fe_{3-\delta}O_4$ bimagnetic NCs (d< 25nm) [34]. The beneficial influence of defective structures at subcritical nanocrystal sizes (d~ 8-10 nm), has also been demonstrated with doped ferrites. In such cases, partial substitution of Fe by transition metals in the nanostructure, optimizes the effective magnetic anisotropy of Iron Oxide Nanocrystals (IONCs). [19] While this pathway maintains the desired morphology/size, it also provides improved heating generation (cf. Specific Absorption Rate, SAR), at least up to a certain level of substitution (e.g. at x~ 0.6, in $Co_xFe_{3-x}O_4$). [35,36]

Keeping in mind these findings, the present work aims to shed light on how atomic scale defect control can be manipulated by chemical substitution in suitably shaped, sub-critical size $Co_yFe_{1-y}O@Co_xFe_{3-x}O_4$ core@shell IONCs. For this purpose, synchrotron X-ray total scattering was utilized to probe their structure evolution at different Co-substitution levels. The findings were complemented by atomistic Monte Carlo simulations, which together with bulk magnetic measurements, offered a means to rationalize how tuning the population of crystal lattice vacancies may tailor heating power generation for hyperthermia applications.

## Experimental

### Synthesis of core@shell $Co_yFe_{1-y}O@Co_xFe_{3-x}O_4$ IONCs

Here we follow a synthetic route designed to produce uniform size, core@shell $Co_yFe_{1-y}O@Co_xFe_{3-x}O_4$ colloidal NCs, which was slightly modified from a previously demonstrated procedure that produced pure $Fe_xO@Fe_3O_4$ particles.[34,37,38,39] Importantly, this allows for direct comparison of the magnetic behaviour and structural characteristics among the series. The chosen protocol offered nanocrystal samples in the size range of 14-18 nm that is below the nominal critical size for IONCs to present superparamagnetic behavior and enhanced magnetic losses, adequate for energy conversion into heat.[28] The protocol enables easy tuning of size and shape (spherical or cubic), while preserving a well-defined core-shell structure and good control of composition; the cobalt content (x,y) of the samples made here varies from 12% to 35%. As mentioned earlier, the substitution of an ion in the host material, could be realized through ion exchange. However, there is a drawback, associated with this technique, when implemented in Wüstite@Spinel-ferrite core@shell-type NCs.[40] Thermal treatment of the pre-formed NCs could compromise the highly sensitive core@shell structure, causing an important oxidation of the core.[41] Even mild heating of the NCs in a colloidal solution containing the ions involved in the ionic exchange procedure, can provide the chemical potential for the initiation of diffusion effects, in both the rock-salt and spinel crystal structures, which lead to the oxidation of the rock-salt phase (core) and the particle conversion to the spinel type.[42]

Here, we aim to suggest ways to improve the magnetic response and especially the magnetically induced heating performance of the MNPs, which is associated with the existence of an interface, between an AFM phase in the core and a FiM phase in the shell, while other types of magnetic anisotropy are also operational (cf. magneto-crystalline, surface anisotropy etc.). Thus, we chose to grow the substituted MNPs either (i) by mixing the desired amounts of two distinct metal-organic precursors, one containing iron and one containing cobalt, at the initial stages of the reaction, a strategy which ensures the preservation of both magnetic phases or (ii) by utilizing a single metal-organic precursor, containing both iron and cobalt in a predetermined ratio. Both approaches produced well-shaped $Co_yFe_{1-y}O@Co_xFe_{3-x}O_4$ NCs, with adjustable Co/Fe levels. The method of mixing different precursors, exclusively containing one metal ion, is slightly more practical, since it allows for a wider Co-content range, just by varying the precursor proportion during the synthesis, whereas the other approach is more favorable when precise control of the Co-content of the nanoparticles is needed.

### Materials

All reagents were used as received without further purification. Oleic acid (technical grade, 90%), octadecene (technical grade, 90%), hexane (ACS reagent, ≥99%), absolute ethanol (≥98%) and sodium oleate powder (82%) were purchased from Sigma Aldrich. Iron (III) chloride ($FeCl_3 \cdot 6H_2O$, ACS/Ph Eur reagent, ≥99%) and Cobalt (II) chloride ($CoCl_2 \cdot 6H_2O$, ACS reagent, 98%) were purchased from Merck. Deionized water was used when needed. Fe(III) oleate, Co(II) oleate and mixed Fe(III)/Co(II) oleate precursors were synthesized in the laboratory, as described in the following protocols.

### Syntheses protocols

Colloidal syntheses were carried out in 100 mL round bottom three-neck flasks connected via reflux condensers to a standard Schlenk line setup. Immersion temperature probes and digitally controlled heating mantles ensured accurate temperature control of the colloidal mixture. Ar gas has been used as a protective atmosphere. Previous studies have shown that $Fe_xO$ NCs formed under such synthetic conditions become oxidized to $Fe_3O_4$ and/or $\gamma$-$Fe_2O_3$ after removing the Ar-blanket and exposing them to ambient air.[43] Alternatively, a 5% $H_2$/Ar gas mixture can also be used as a protective atmosphere in order to take advantage of the reductive action of $H_2$, to further protect the sensitive rock-salt from being oxidized to the spinel forms. Since Co and Fe are introduced simultaneously into the colloidal mixture, it is believed that, based on the shell formation mechanism, Co is equally distributed in the whole nanoparticle volume, namely in both the core and the shell phases.

**Preparation of metal-oleate precursor.** A mixed Co(II)/Fe(III) – oleate, with the desired Co/Fe ratio, was prepared before each nanoparticle synthesis and subsequently used as iron and cobalt precursor. The metal oleate precursor was formed by the decomposition of $FeCl_3 \cdot 6H_2O$ and $CoCl_2 \cdot 6H_2O$ in the presence of sodium oleate at ~60$^oC$, based on our previous protocol for Fe-oleate formation.[34,44] In a typical synthesis, 16 mmol of $FeCl_3 \cdot 6H_2O$ salt and 48 mmol of sodium oleate were dissolved in a mixture of solvents, namely 56 mL hexane, 32 mL ethanol and 24 mL deionized water, in a round bottom three-neck flask. The required amount of $CoCl_2 \cdot 6H_2O$ was added in the mixture, adjusting accordingly the sodium oleate quantity. We were doing so, considering that for the stabilization of each Fe(III), three oleate anions and for each Co(II), two oleate anions are needed. We were though keeping the solvent quantities the same. The mixture was heated to 60-65$^oC$ under Ar atmosphere for 4 hours and then left to cool down to room temperature. A separatory funnel was then used to separate the upper, organic phase, containing the metal oleate complex, from the aqueous phase. Afterwards, the organic phase was washed with about 30 mL deionized water and separated again, a process which was repeated 4 times. At the end, the metal organic complex was dried under stirring and mild heating for several hours, resulting in a viscous metal-oleate, with its color ranging from dark red to brownish, depending on its Co content. Special care was taken to protect the final product from the light. Some mild heating to ensure its fluidness may be needed just before its use for each nanoparticle synthesis. A modified synthetic route for the fabrication of nanoparticles was also tested, where two distinct metal oleate precursors were used to supply the system with Fe(III) an Co(III) species. Pure Fe(III) – oleate was formed following exactly the same protocol, omitting the addition of Co(II)[34] and then, pure Co(II) – oleate was similarly produced by dissolving 16 mmol $CoCl_2 \cdot 6H_2O$ and 32 mmol sodium oleate in 42 mL hexane, 24 mL ethanol and 18 mL deionized water. The oleates were separated, washed and dried in the same way as described above and the desired amount of each one was then used before each nanoparticle synthesis. We found both ways to be sufficient for the production of high-quality nanoparticles.

**Synthesis of Co-substituted iron-oxide nanocrystals.** The nanoparticles were synthesized by employing a protocol similar to our earlier work[34] that obeyed a slightly modified previous synthesis avenue.[37,38,39] In a typical synthesis 2.4 mmol of mixed iron/cobalt oleate or 2.4 mmol in total of iron oleate and cobalt oleate were dissolved in octadecene. The proportion of Fe/Co was calculated and predetermined according to the desired ratio of metals in the final product. 1.2 mmol of oleic acid was added as surfactant. The amount of the solvent (octadecene) was tuned so that a final Fe/Co-oleate molar ratio of 0.2 mol/kg solution was achieved. Three discrete major steps can be identified during the synthetic protocol. First, the mixture is being heated at 100 $^oC$ under vacuum for 60 min at a degassing step, for the complete removal of any water and oxygen residues. This is crucial and possible omission of this stage may lead to single-phase ferrite particles.[45] Then, the mixture is heated to 220 $^oC$, with a heating rate of about 10 $^oC$/min. This is the so-called "nucleation step", which allows for the crystal seeds to be formed. At the final step, known as the "growth step", the colloidal mixture is heated to 320 $^oC$ with the same heating rate, where the nanocrystals' growth takes place. The effective separation of the nucleation and growth steps is crucial for the production of monodispersed nanoparticles. Minor variations in this two-step heating protocol ensure the control of the particles' size distribution and allow the tuning of their size. The latter is mainly achieved through the time of stay at each of the 2 final steps. Our protocol, with a time of stay of about 60 min at each of these steps, gives rise to spherical NPs with average diameters in the 14 – 18 nm range. As proposed in earlier studies, the addition of sodium oleate powder in a proportion of 1:8 – 1:5 with respect to the metal-oleate precursor, is adequate to promote the formation of cubic NPs.[39] Under the described experimental conditions, slightly larger nano-cubes, with edge diameters ranging from 18 to 20nm can be realized this way. Though, the mixing of two separate oleate precursors, or the use of a mixed Fe/Co oleate precursor, seems to make the fine tuning of the Sodium oleate/metal oleate ratio a bit trickier, for the fabrication of well-formed cubic NPs, compared to pure iron oxide NPs. At the end of the synthesis the colloidal mixture, containing the nanocrystals (NCs), was left to cool down at room temperature. The NCs were precipitated upon ethanol addition. They were then separated by centrifugation at 6000 rpm for 5 min, re-dispersed in hexane and centrifuged once more after adding ethanol in a 1 : 1 ratio with respect to the hexane. A short stay in a sonication bath may have be needed after the centrifugation for the complete re-dispersion of the particles. Excessive sonication though might affect the quality of the dispersion, since this removes the surfactants from the particle surface. The whole process was repeated two more times at a centrifugation speed of 1000 rpm. Although the size distribution is tuned during the synthesis, a well-chosen post-synthetic purification protocol is essential for an effective size-selective separation of the MNPs.[46]

## Characterization techniques

### Morphological Characterization

**High resolution transmission electron microscopy.** High resolution transmission electron microscopy was utilized to reveal the morphological characteristics of the prepared nanoparticles, namely the size, shape and core@shell structure. Low-magnification and HRTEM images were recorded using a $LaB_6$ JEOL 2100 transmission electron microscope, operating at an accelerating voltage of 200kV. A Gatan ORIUS™ SC 1000 CCD camera was used to capture the images. The average size of the particles arises from a statistical analysis, measuring the diameter or edge length of a statistically acceptable amount of nanoparticles with the free image processing software *ImageJ*.[47] The nanoparticle suspensions needed to be highly diluted before the TEM analysis. The same solvent, hexane, was utilized for such a dilution. A drop of the diluted nanoparticle suspension was then deposited onto a carbon-coated copper grid, allowing the hexane dispersant to evaporate. The image is formed from the interaction of the electrons with the nanoparticles as the electron beam is transmitted through the specimen. The TEM column is evacuated to a pressure down to the order of $10^{-4}$ Pa to reduce the chance of interaction between electrons and gas molecules. The beam is focused when passing through the electromagnetic lenses in the electron column and creates a magnified image on a fluorescent screen, making the observation by the operator possible. The screen can be retracted to allow the CCD camera located below to record the image. Contrast can arise from differences in density. TEMs in high resolution mode are capable of achieving atomic-scale resolution.

### Structural Characterization

**X-ray pair distribution function.** Synchrotron X-ray total scattering data were acquired at the 28-ID-2 beamline of the National Synchrotron Light Source II (NSLS-II), at Brookhaven National Laboratory. Nanoparticle dispersions were dried (hexane was evaporated under ambient conditions) and the resulting nanoparticle powder was encapsulated in Ø1.0 mm Kapton capillary, sealed at both ends with epoxy glue. The beamline, was setup in "PDF" data collection mode, with a Perkin-Elmer 2D image plate detector (sample-to-detector distance of 326 mm, calibrated against Ni standard) for fast data acquisition, but of relatively modest $Q$ space resolution, which limits the PDF field of view in $r$ space.[48] Data were collected on warming between 10 K and 300 K, in 5 K steps, making use of the beamline's continuous flow liquid helium cryostat (Cryo Industries of America), with incident X-ray energy of 68 keV (λ= 0.1823 Å). Data for bulk cobalt ferrite ($CoFe_2O_4$) and wüstite ($Fe_xO$) powders were also acquired as a reference.

The atomic pair distribution function (PDF) gives information about the number of atoms in a spherical shell of unit thickness at a distance r from a reference atom[49] and is defined as

$$\boldsymbol{G}(\boldsymbol{r}) = 4\pi\, \boldsymbol{r}[\boldsymbol{\rho}(\boldsymbol{r}) - \boldsymbol{\rho_0}] \tag{1}$$

where $\rho_0$ is the average number density, $\rho(r)$ is the atomic pair-density, and $r$ represents radial distance. The raw 2D experimental data are then converted to 1D patterns of intensity versus momentum transfer, $Q$, utilizing the Fit2D program,[50] which are further reduced and corrected using standard protocols including the use of PDFgetX3 software,[51] and then finally Fourier transformed to obtain G(r):

$$G(r) = (2/\pi) \int_{Q_{\min}}^{Q_{\max}} Q\,[S(Q) - 1]\,\sin(Q\,r)dQ \tag{2}$$

*Q* (4π sinθ/λ) is the magnitude of the momentum transfer for elastic scattering, and *S(Q)* is the properly corrected and normalized powder diffraction intensity measured from $Q_{min}$ to $Q_{max}$ (0.25≤ *Q*≤ 25 $Å^{-1}$). The PDF data, collected over a sufficiently wide Q-range, thus carry structural information over a broad range of length scales, in contrast to traditionally used (bulk) probes, like EXAFS and NMR, which are dominated by the nearest-neighbor interactions. The experimental PDF, *G(r)*, can subsequently be modelled by calculating the following quantity directly from a presumed structural model:

$$G(r) = \left[\frac{1}{r}\sum_{\mathrm{ij}}\frac{f_i f_j}{<f>^2}\,\delta\left(r - r_{\mathrm{ij}}\right)\right] - 4\pi\, r\, \rho_0 \tag{3}$$

Here, *f* stands for the X-ray atomic form factors evaluated at *Q= 0*, $r_{ij}$ is the distance separating the i-th and j-th atoms, and the sums are over all the atoms in the sample. In the present experiments, elemental nickel powder was measured as the standard material to determine parameters, such as $Q_{damp}$ and $Q_{broad}$, needed to account for the instrument resolution effects. Raw data (for a $Q_{max}$= 25 $Å^{-1}$) were fitted with the PDFgui software suite.[52]

**Energy-dispersive X-ray spectroscopy (EDS).** Elemental analysis of the as prepared samples was performed with a JEOL JSM-6390VL Scanning Electron Microscope (SEM), operating at 20kV, equipped with an INCA PentaFET-x3 EDS analyzer from Oxford Instruments. X-rays are produced when a beam of electrons, coming from the electron gun of the microscope, stimulates the sample, which is located on a movable stage in the sample chamber. When ground state electrons in discrete energy levels are excited, creating electron holes, electrons from higher-energy electron shells fill these holes, emitting X-rays, which correspond to the energy difference between the higher- and lower-energy levels. Each chemical element has a unique atomic structure, which leads to an X-ray emission pattern that is characteristic for the specific electron configuration. The number and energy of the X-rays emitted from a specimen are being measured by the energy-dispersive spectrometer, allowing for compositional analysis of the sample. EDS can determine the elements that are present in a sample and their relative abundance in it. The accuracy of the method though, can be affected by the nature of the sample and the possible overlap of X-ray emission peaks from elements with similar electron configurations.

**Magnetic Characterization**

The magnetic behaviour of the nanoparticle samples was evaluated using a vibrating sample magnetometer (VSM, Oxford instruments, Maglab 9T), operating at a vibration frequency of 55 Hz. The measurements of the temperature-dependent magnetization, *M(T)*, were carried out at 50 Oe at a fixed temperature rate of 1 K/min after either zero-field cooling (ZFC) or field cooling (FC) in 50 Oe from 5-300 K. Hysteresis loops, *M(H)*, were obtained at room temperature and 5K by sweeping the applied field from +50 kOe to -50 kOe and back to +50 kOe after cooling the sample from 300 K to 5 K under ZFC or an applied field 0 < $H_{cool}$ ≤ 50 kOe (FC) for selected samples. In the FC procedure, once the measuring temperature was reached, the field was increased from $H_{cool}$ to *H* = 50 kOe and the measurement of the loop was pursued. The *M(H)* data were recorded under a magnetic field sweep with an optimised rate of 30 Oe/s in order to minimize the propagation of a possible synchronization error of the measuring electronics, present in the case of sweeping rates greater than 200 Oe/s. Special precautions were taken to maintain the structural integrity of the samples and more specifically to avoid a possible spontaneous oxidation of the rock-salt core to spinel. The specimens for magnetic characterization were prepared by absorbing colloidal

MNPs dispersions on pre-weighed cotton wool pieces and letting the hexane to evaporate, until a pure nano-powder weight of ~10 mg was obtained. They were subsequently placed in suitable size gel capsules, and then sealed under Ar atmosphere (in a Glovebox) in airtight vials. The vials were then opened just before the magnetic measurements to avoid unnecessary exposure to ambient air. This ensured that the magnetic behavior of the samples matched the structure probed by the synchrotron X-ray studies, as similar actions were undertaken at the NSLS-II facility.

**Monte Carlo Simulations**

In our effort to interpret the experimental results for the MNPs we have developed a microscopic core/shell model, where the interface (IF) of the two regions was explicitly included. The calculations were carried out with the Monte Carlo (MC) technique and the implementation of the Metropolis algorithm.[53] Spherical MNPs were considered, with a diameter d, expressed in lattice spacings of a simple cubic lattice (cf. magnetite cell, a= 8.39 Å). The simulated MNPs consist of an AFM core and a FiM shell.[54] For this purpose, we studied three model systems, with different structural features, to simulate the magnetic behaviour of a pure core@shell particle ("pure" corresponds to $Fe_xO@Fe_{3-\delta}O_4$ MNPs, identified as sample S15 in reference[34]), and two Co-substituted core@shell MNPs involving a lower Co:Fe proportion ("model#1") and another one with higher Co:Fe content ("model#2"). The radius of these three modelled MNPs remains the same, while other structural characteristics (e.g. core/shell volume ratio, vacant sites etc.) differentiate, according to the degree of Co-substitution at the Fe-sites. Such a parametrization was motivated by the trends derived from the experimental xPDF data and the subsequent structural analysis. The exact morphological and structural characteristics of the simulated model MNPs are schematically illustrated in **Fig. 1**, and are as follows:

**"pure"**, spherical MNPs of an average radius R=9.1 and a shell thickness of 4 lattice spacings, assume 20% core and 80% shell, as a fractions of the whole particle volume, entailing an AFM core and a FiM shell. The MNPs are assumed to have 25% vacant metal-ion crystallographic sites (with no cobalt content), randomly dispersed in *both* the core and shell phases.

Introduction of Co in the crystal structure at different levels, assumes two $Co_yFe_{1-y}O@Co_xFe_{3-x}O_4$ MNPs of the same average radius, namely:

**"model#1"** , the MNPs have a radius of R=9.1, shell thickness of 3.3 lattice spacings, resulting in 25% core and 75% shell volume, with a Co substitution level of 10% in both phases, meaning that now 10% of the provided Fe-sites in the core are occupied by Co and 10% of Fe-sites in shell are also occupied by Co. The model adopts 40% vacant metal-ion sites, randomly distributed over the core phase only, including the core interface. (see eq. 4)

**"model#2"** , the MNPs have a radius of R=9.1, shell thickness of 1.9 lattice spacings, thus resulting in 50% core and 50% shell volume, with a Co substitution level of 35% in both phases, following the same concept as in model#1. This model adopts 40% vacant metal-ion sites, randomly distributed over the shell phase only, including the shell interface and surface.

The spins in the MNPs were assumed to interact with nearest neighbor Heisenberg exchange interaction and at each crystal site they experience a uniaxial anisotropy. Under an external magnetic field, the energy of the system is calculated as:[53,54,34]

$$E = -J_{core}\sum_{i,j\in core}\vec{S}_i\cdot\vec{S}_j - J_{shell}\sum_{i,j\in shell}\vec{S}_i\cdot\vec{S}_j - J_{IF}\sum_{i\in core, j\in shell}\vec{S}_i\cdot\vec{S}_j$$

$$-K_{i\in core}\sum_{i\in core}\left(\vec{S}_i\cdot\hat{e}_i\right)^2 - K_{i\in shell}\sum_{i\in shell}\left(\vec{S}_i\cdot\hat{e}_i\right)^2 - \vec{H}\sum_{i}\vec{S}_i \qquad (4)$$

Here $\vec{S}_i$ is the atomic spin at site *i* and $\hat{e}_i$ is the unit vector in the direction of the easy-axis at site *i*. We consider the magnitude of the atomic spins in the two AFM sublattices equal to 1 and in the two FiM sublattices of the shell to be equal to 1 and 1.5, respectively. The first term in eq. 4 gives the exchange interaction between the spins in the AFM core and the second term gives the exchange interaction between the spins in the FiM shell. To take into account the difference of the magnetic transition temperatures [$T_N$(core) <$T_C$(shell)], we consider the exchange coupling constant of the core as $J_{core}$ = -0.1 $J_{FM}$ and that of the shell as $J_{shell}$ = -1.5 $J_{FM}$, where $J_{FM}$ is a reference value and it is considered to be the exchange coupling constant of a pure ferromagnet (FM), $J_{FM}$= 1. The third term gives the exchange interaction at the interface between the core and the shell. The interface includes the last layer of the AFM core and the first layer of the FiM shell. The exchange coupling constant of the interface $J_{IF}$ is taken $J_{IF}$= -0.3 $J_{FM}$ . The fourth term gives the anisotropy energy of the AFM core, $K_C$. If the site *i* lies in the outer layer of the AFM core $K_{i\text{-core}}$ = $K_{IF}$ = 5 $J_{FM}$ (due to strong lattice mismatch) and $K_{i\text{-core}}$ = $K_C$ = 0.5 $J_{FM}$ elsewhere. The fifth term gives the anisotropy energy of the FiM shell, which is taken as $K_{shell}$ = 0.1 $J_{FM}$ and at the shell IF $K_{IF}$ = 5 $J_{FM}$. If the site *i* lies in the outer layer (i.e., the surface) of the shell then the anisotropy is taken as $K_{i\text{-shell}}$= $K_S$= 1.0 $J_{FM}$, which is assumed to be random (rather than uniaxial). The last term in eq. 4 is the Zeeman energy. The relative values of $J_{shell}$ and $K_S$ were calculated starting from the bulk values and then taking into account the size of the experimentally studied MNPs and the corresponding surface effects, in line with earlier atomistic simulations.[34],[53],[55] Moreover, in the three models depicted **Fig. 1**, the structural defects (vacancies) were also taken into account as described by Lappas et al.,[34] while in models #1 and #2, the Co ions were randomly distributed, assuming randomly oriented anisotropy axes, with $K$= 5 $J_{FM}$ (10 times larger than the $K_{core}$), since the magnetocrystalline site anisotropy of Co ions is ten times larger than that of Fe ions.[56]

On these grounds, the hysteresis loops *M(H)* were calculated upon a field-cooling procedure, starting at a temperature T = 3.0 $J_{FM}/k_B$ and down to $T_f$ = 0.01 $J_{FM}/k_B$, at a constant rate under a static magnetic field $H_{cool}$, directed along the z axis. The exchange bias field was estimated by the hysteresis loop shift along the field axis $H_{EB} = -(H_{right} + H_{left})/2$ and the coercive field was defined as $Hc = (H_{right} - H_{left})/2$, where $H_{right}$ and $H_{left}$ are the points where the loop intersects the field axis. The fields *H, Hc,* and $H_{EB}$ are given in dimensionless units of $J_{FM}/g\mu_B$, the temperature *T* in units $J_{FM}/k_B$ and the anisotropy constants *K* in units $J_{FM}$. In this work, $10^4$ MC steps per spin (MCSS) were used at each field step for the hysteresis loops and the results were averaged over 60 different samples.

## Results and discussion

### Cobalt-substituted spinel ferrites

In the spinel $(M^{3+})_8[M^{3+},M^{2+}]_{16}O_{32}$ (M= transition metal) structural type [where round brackets represent tetrahedral (Td) and the square brackets octahedral (Oh) coordination by oxygen crystallographic sites], typically, $Co^{2+}$ [57] tends to replace $Fe^{3+}/Fe^{2+}$ resting at the Oh sites [58] although some distribution at the tetrahedral spinel sites (cf. Td) cannot be excluded (Fig. 2). In such ferrospinels, the ratio of $Co^{2+} \rightarrow Fe^{3+}$/ $Co^{2+} \rightarrow Fe^{2+}$ also affects the degree of inversion,[18] but based on the characterization tools at our disposal (i.e. synchrotron X-ray PDF is not able to distinguish the different metal species) we cannot conclude on such effects. As the substituting ion is introduced in the mixture of reactants simultaneously with the other precursors, at the early stages of the reaction, we are of the opinion that $Co^{2+}$ ions are distributed over the whole volume of each particle, in both core and shell, proportionally to their relative volume fraction. This behavior would be in line with earlier findings for the partial oxidative conversion of the initially grown wüstite to create the spinel shell.[59,60,42,18,61] High resolution single-particle elemental mapping of the MNPs would be required to confirm this proposition.

**Structural insights from microscopy**

**HRTEM: Single-particle morphology.** Transmission Electron Microscopy is the basic, but important post-synthetic characterization tool to evaluate the success of the synthetic procedure. Good dispersion quality (no agglomeration of MNPs), narrow size distribution and well-defined particle shapes are the key factors to look for. Samples that did not meet these criteria were rejected. Thereafter, three Co-substituted nanoparticle samples of spherical shape, in the 14-18 nm size range, were chosen for the study. Specifically, they attained diameters of d= 15.2 ± 1.2 nm, d= 13.9 ± 1.1 nm and d= 17.8 ± 0.9 nm, with various Co-substitution levels of 12%, 21% and 35%, respectively. The ability to change the particle morphology from spherical to cubic, by slightly changing the amount of surfactants, has been also confirmed (ESI, Figure S3). Cubic samples though are not thoroughly discussed in the following sections, due to lack of a more systematic variation in their Co-substitution rate. Since the Co content seems to be the dominant effect that defines their structure and magnetic response, these samples are called henceforth S12, S21, S35 respectively, where S stands for spherical morphology, while the following numbers are indicative of the cobalt substitution level (see following SEM-EDS section). HRTEM (**Fig. 3**and **Figure S3-a**) suggests that all nanoparticle samples show a high degree of monodispersity, in terms of shape and size (narrow size distributions). The coexistence of dark and light contrast features in all samples (clearly shown in high-magnification TEM images) is a proof that two chemical phases of different electron diffracting power share the same nanocrystal volume. Since the fabrication of the specimens followed a similar synthetic route to the one that yielded $FeO@Fe_3O_4$ NCs studied before,[34] the aforementioned findings are in line with the tendency of such NCs to adopt a topological arrangement of phases, with rock salt type in the core of the particle that is surrounded by a spinel type of phase in the shell (*vide infra*: x-ray PDF analysis).

**SEM-EDS: Compositional analysis.** Energy-dispersive X-ray spectroscopy was used to estimate the average Co/Fe ratio from an ensemble of nanoparticles. The specimens for the EDS analysis were prepared by casting a few drops of the as-prepared colloidal suspension on a silicon substrate and then letting the dispersant to evaporate. They were then placed on the specialized stage of the SEM sample chamber, pumped for a few minutes to achieve the required vacuum and finally the electron beam was focused on the specimens using the special electromagnetic lenses located in the electron column. The measurements were taken from several sections of the specimens by focusing the beam on different spots of the specimen surface, for statistical reasons. This procedure suggests a high compositional uniformity across the surface of the specimen. The reported results represent averages of the different measurements. Fortuitously, iron shows a strong and distinct characteristic peak, which does not overlap with the cobalt peaks, allowing for a reasonably accurate calculation of their relative abundance from the relative peak intensity, compared to measured standard samples. The spectra show only the characteristic peaks attributed to Fe and Co, apart from a strong silicon peak coming from the substrate, indicating high chemical purity for all samples (**ESI, Figure S4**). The relative atomic composition in all cases agrees within the accuracy of the assessment with the nominal one, planned during the chemical synthesis. It is worth noting that the Co substitution level refers to the overall Fe ions being replaced by a certain amount of Co. The SEM-EDS facility utilized here though, cannot derive the exact Co-level in the individual core and the shell topological sections. For the following analysis, we assume that Co-substitution is equally possible in both crystallographic phases and the probability of Fe ion to have been replaced by Co in each section is proportional to their relative volumetric ratios.

**Structural insights from synchrotron X-ray PDF**

The need for both qualitative and semi-quantitative, phase-specific structural information from large ensembles of particles that could reveal Angstrom-long localized lattice and bonding distortions in the unit cells, motivated us to acquire the synchrotron X-ray total scattering data, for all nanoparticle specimens and compare them to the bulk reference materials (**Fig. 4**). These, combined with the PDF method offer insights that move beyond the findings of techniques that focus on single nanoscale particles (e.g. HRTEM).[40] Our effort here is concerned with (i) confirming the coexistence of two distinct crystallographic phases (i.e. rock salt and spinel) in a core@shell structure, (ii) estimating their average relative volume fractions, (iii) unveiling possible deviations of the local structure from cubic symmetry and (iv) uncovering likely temperature-mediated effects; thus, we recorded the PDFs for all the materials between $10 < T < 300$ K. Our analysis addressed mainly the low-r region in the atomic PDFs (r= 1-10 Å), as the field of view was limited due to the moderate Q space resolution of the experimental setup used.

**Local structure – rock salt phase modifications**

Bulk $CoFe_2O_4$ was measured as a reference material. However, due to the similar electron count for Fe and Co, there is lack of sensitivity in X-ray PDF (xPDF) to differentiate between the two. For this reason, without loss of relevant information to which the method is sensitive to we have utilized models approximated by the magnetite $Fe_3O_4$ composition. The technique though is able to identify the differences between tetrahedral (Td) and octahedral (Oh) crystallographic sites, based on their quite different chemical environments. Thus, an iron-only spinel offers an adequate approximation for the description of our systems. This is confirmed by the observation that the atomic PDF in the low-r region for the bulk reference sample ($CoFe_2O_4$) is described equally well with the cubic cobalt ferrite spinel, as well as with the cubic magnetite models (**Figure S5, in ESI**). As the valence state differences are indistinguishable too, a possible site inversion could not be verified and the degree of inversion also could not be determined with this technique. Therefore, we performed the structure refinements based on the simplified, normal spinel, cubic configuration of magnetite, $(Fe^{3+})_8[Fe^{3+},Fe^{2+}]_{16}O_{32}$.[62] Especially in the case of our model, this could be simply expressed as $(M)_8[M]_{16}O_{32}$, where M stands for transition metal cation (Fe or Co), the round brackets represent tetrahedral (Td) and the square brackets octahedral [Oh] coordination by oxygen crystallographic sites, assuming no $M^{2+}/M^{3+}$ inversion between Td and Oh sites. This model was then utilized to fit the xPDF data for nanoparticles of variable Co-concentrations, with a reasonably good quality of the fit ($R_w$) in the low-r region, at room temperature. However, the introduction of an additional crystallographic phase of rock-salt type cubic cell ($Fe_xO$)[63] is essential (*vide infra*) for obtaining adequate fit quality (**Fig. 5**).

The resulting simplified 2-phase rock-salt@spinel cubic model describes well all the nanoparticle samples, unlike our previous work,[34] where this model systematically failed to describe the peak at r ~ 3 Å (PDF data of $Fe_xO@Fe_{3-\delta}O_4$ or the fully oxidized $Fe_{3-\delta}O_4$ NCs), which implied a deviation of the nanoparticle local structure from the ideal cubic lattice configuration, that was better described with a distorted, tetragonal spinel model.[34] In the samples studied here, there is no evidence for a local structure symmetry lowering from cubic to tetragonal. Another significant difference in these samples, arises from their tendency to preserve the core@shell structure, as shown in the fit results, from the increased volume fraction of the rock-salt phase. (**see ESI, Table S3**) Interestingly, the higher the Co-content, the larger the volume fraction of the rock-salt phase. More specifically, as evidenced from the raw data, the nanoparticle PDF peak centered at r~ 3 Å shows a systematic shift to higher interatomic distances (of the same extent for all samples), compared to bulk spinel reference (**Fig. 6**, **Fig. 7**). At the same time, there is a clear evolution in peak intensity as Co content rises (**Fig. 7**). It is helpful though to refer to the PDF peak identification, based on the expected positions of metal (Fe) interatomic distances in a 2-phase rock-salt@spinel system, as shown in **Fig. 7** in the form of calculated partial PDFs. The peak at r~ 3 Å may correspond to the closest distance between the metal atoms at the Oh sites in the spinel structure (**Fig. 2**, **Fig. 7**) Nevertheless, the experimentally observed shift to higher-*r* can be attributed to the strong presence of rock-salt phase in all samples, since the nearest distance of a pair of M-M in the rock-salt is just above 3 Å (see **Fig. 7**, green line). That is why we were able to account for it when fitting the G(r) data by introducing an additional phase of rock-salt type in the model. Along these lines, the observable increase in peak intensity can also be justified by the growth of the rock-salt volume in the MNPs, occurring during the increase of Co content, as demonstrated by the PDF fit results (**ESI**, **Table S3**).

Although all fits were of reasonably good quality, significant conclusions were made from model-independent assessment of the raw PDF data and related simulations, compared to the PDF data obtained for the $Fe_xO@Fe_{3-\delta}O_4$ nanoparticle system reported before.[34] In the latter, the IONCs displayed variable shifts of the PDF peaks at r ~ 3 Å (attributed to $Fe_{Oh}$-$Fe_{Oh}$ and $Fe_{Oh}$-$Fe_{Td}$ interatomic distances, **see Fig. 2, Fig. 7**, purple and red lines), even when no rock-salt phase was present, implying deviations from the ideal cubic local structure, stabilized by variable levels of Fe-site vacancies present in the spinel phase.[34] In the Co-substituted samples, though, the peak shift towards higher r-values is accompanied by a dramatic increase of the peak intensity, clearly originating from rock-salt phase contribution, as corroborated by our PDF simulations (**Fig. 8 a,b**). More specifically, the simulation shown in **Fig. 8-b** implies that in a two-phase model, if a rock salt contribution is enhanced (no other changes made in the model) it causes the 3 Å peak intensity to grow and the 3.5 Å intensity to drop. Notably, the higher the Co-content in the nanoparticle samples, the higher the peak associated with the rock-salt phase, as presented in **Fig. 8-a**. This is in line with the findings from data fittings over the low-r region (1 nm) of the PDFs. Thus, it seems that chemical substitution of Fe for M (M = Mn, Co, Ni), has two effects: it does not only partially eliminate the local distortions observed in pure iron oxide specimens[64], but also helps the preservation of the rock-salt core, at least in this type of Co-substituted core@shell systems. This could be rationalized by a less defected structure of the rock-salt cobalt-based oxide (CoO) core, while the $Co_xFe_{3-x}O_4$ shell supports a higher resistance to oxygen diffusion.[61]

The behavior may imply that upon creation, the Co-mediated spinel shell acts as an effective barrier, impeding further core oxidation, commonly observed in heterostructured $FeO@Fe_3O_4$ nanoparticles.[40] Indeed, earlier studies found that 12.5 nm, Co-substituted FeO presents limited surface oxidation, with a rock-salt@spinel structure, where the core is preserved 120 days after being exposed to air,[61] while even smaller, 9.5 nm, MNPs appear not to change with aging, as indicated by their rather unchanged magnetic behaviour, 120 days after exposure to ambient conditions.[65] The observation suggests that samples richer in Co exhibit a better-preserved core, possibly arising from the enhanced stability of the $Co^{2+}$ ions against oxidation to $Co^{3+}$ rather than that of $Fe^{2+}$ counterparts, hence paving the way for synthesizing core@shell nanostructures that are highly oxidation-resistant over time. The observed intensity increase of the PDF peak in question, also complies with a higher metal ion occupancy in the rock-salt phase, due to its less defective structure (**Fig. 2**).

**Local structure – defect mediated changes**

When contribution of the rock-salt phase is enhanced, as described above, an inevitable lowering of the peak maximum at r ~ 3.5 Å is to be expected, changing the 3.0/3.5 Å relative peak intensity ratio, favoring the 3 Å peak. This phenomenon is observed in the PDF data, but it is additionally intensified, implying possible sub-stoichiometry effects, which further lower the intensity of the 3.5 Å peak. The peak at r ~ 3.5 Å corresponds to the closest distance between $M_{Oh}$ and $M_{Td}$ in the spinel lattice, **(Fig. 2)** while a weaker contribution attributed to $M_{Td}$ - $M_{Td}$ pairs, at slightly higher interatomic distances (*r*) affects this peak as well (**Fig. 7**), as M ions could be either Fe or Co in the spinel structure in our nanoparticle samples. This complexity makes it challenging to investigate whether the peak intensity suppression is caused by atomic site defects (i.e. vacant crystallographic sites) either in $M_{Oh}$ sites alone, or $M_{Td}$ sites alone, or both. To further explore whether the dramatic changes in the relative PDF peak intensities are indeed driven by the presence of vacancies at M sites, and to evaluate whether these vacancies have any site-specific preference, simulated xPDF patterns were obtained based on the normal cubic spinel configuration, while the population of either M-vacancies at the Oh sites or M-vacancies at the Td sites was varied (**Fig. 8 c,d**). The simulations show that the higher the level of $M_{Oh}$ vacancies is in the spinel structure, the less intense the 3 Å peak becomes, whereas the 3.5 Å peak is rather unaffected (**Fig. 8-c**). Additionally, when $M_{Td}$ sites become vacant, this decreases dramatically the peak intensity at 3.5 Å and somewhat increases the intensity of the 3 Å peak (**Fig. 8-d**). Combining the above-described trends, we conclude that given a significant percentage of rock-salt volume fraction, M sub-stoichiometry at the Td sites is the dominant effect, although some extent of vacancies on Oh sites cannot be excluded.

To independently verify this proposal, we further extracted the Atomic Displacement Parameters (ADPs) as a function of temperature (T= 10-300 K), based on the 2-phase model fits of the PDF data mentioned earlier. ADPs reflect atomic thermal motion and possible static disorder of atoms in the structure.[66] Isotropic T-dependent ADPs (characterised by $U_{iso}$/Å$^2$) for M sites in the two phases considered, are plotted in **Fig. 9**. All the ADPs in sample S12 show similar, values that are typical for nanostructured systems,[67,68] displaying a smooth upward trend with increased temperature. When going to Co-richer samples (S21, S35) the ADPs for the Oh metal atoms in the spinel phase and those for the metal atoms in the rock-salt phase, remain approximately the same, as depicted by interpolating the ADP curves to T= 0 K and T= 300 K. On the other hand, the ADPs for the Td metal atoms in spinel grow with Co-content, indicating possibly higher static disorder in the $M_{Td}$ sublattice. The absence of $M_{Td}$ ions from their predicted crystallographic sites, as suggested from the previous analysis based on simulations, makes the corresponding $M_{Td}$-$M_{Td}$ pair bonds more compressible, thus allowing for broader atomic thermal motion. In fact, the $ADP_{Td}$ curve, observably pushes away from the $ADP_{Oh}$ curve on going from S12 to S21 and S35 (**Fig. 9**). This apparent "splitting" becomes more significant as the Co content increases, implying significantly different static disorder associated with more dramatically perturbed chemical environment of the Td coordinated metal ions as compared to the Oh sites. This trend is further corroborated by the extracted Einstein temperature, $\Theta_E$, obtained from fitting the T-evolution of the ADPs with the correlated Einstein-model (**Fig. 9**)[69]:

$$\sigma^2(T) = \sigma_0^2 + \frac{\hbar^2}{2\mu k_B \Theta_E} \coth\frac{\Theta_E}{2T} \quad (5)$$

In this description, assuming Fe-Fe atom pairs only, $\Theta_E$ become significantly smaller for Td ions, when Co-content increases, thus suggesting lattice softening. At the same time a significant increase of $\sigma_0$ for Td sites indicates a notable increase of static disorder in the $M_{Td}$ sub-lattice (**ESI, Table S3**). A similar trend could be recognized for the rock-salt ADPs, which however show slightly lower values upon Co-level increase, implying a subtle stabilization of the rock-salt phase against the spinel one. Minimal variations in the crystallographic site preference of Co ions (Td or Oh) from sample to sample, cannot be excluded as well. Overall, the significant changes in ADPs, $\Theta_E$ and $\sigma_0$ for Td sites confirm that disorder in the Td spinel sites, mediated by Co incorporation in the lattice, is the predominant effect.

A plausible explanation of the above-mentioned effects may arise from an earlier proposed pathway for rock-salt self-passivation, through clusters of defect-units that coalesce in order to nucleate ferrospinel phases.[42,59,60] Based on this mechanism, highly defected rock-salt structures would have a large population of vacant $M^{2+}$ sites in the rock-salt lattice, which would give rise to a significant amount of interstitial $M^{3+}$ ions for charge balancing reasons (**Fig. 2 d**). Such interstitial ions are coordinated by ordered vacant sites (*V*) in such a manner (cf. $V_4$-Td defect clusters),[59] which resembles the Td M sites in spinels (**Fig. 2 e**). Thus, during the rock-salt to spinel oxidative conversion, these interstitials may have the tendency to become M-ions occupying Td atomic sites in the spinel structure. In a less-defective rock-salt system, such as that obtained by Co-substitutions (**Fig. 2 c**), there is less need for such charge balancing, thus resulting in fewer interstitial ions. It seems that the lower population of these interstitials in the Co-mediated rock salt phase, has as a consequence, a lower population of $M_{Td}$ ions in its oxidized form (spinel) that reflects in more abundant Td vacant sites.

In summary, the PDF analysis was able to evaluate the crystal structure of the rock salt core and the spinel shell and quantify their average relative volumetric ratio (**see ESI**, **Table S3**) for each particle batch. The analysis suggests that the incorporation of Co ions in the $Fe_xO$ phase, cures the commonly suffering from defects structure by effectively increasing the occupancy of metal ion sites in the rock salt phase. This seems to give the core a higher degree of resistance against oxidative conversion to a spinel shell. Thus particles with higher Co content have also higher volume fractions of rock salt phase in the core (**ESI**, **Table S3**). The xPDF results indicate that a higher $M^{2+}$ stoichiometry in the rock salt part by Co-substitutions, results in increased levels of unoccupied Td sites in the spinel, likely through oxidative shell creation involving an ordered defect clustering mechanism, while Oh sites may retain a typical moderate sub-stoichiometry, as implied by the comparable across the samples ADPs (**Fig. 9**).

## Magnetic behavior

In view of the variation of Co% content among the nanoparticle samples, the consequent structural evolution of the core@shell structure and the deviation from a perfect, fully stoichiometric crystallographic ordering (*vide supra*), we measured and evaluated their magnetic properties. Our main interest was to investigate whether and at what extent the identified structural characteristics affect the magnetization of the particles, the exchange interactions and the magnetocrystalline anisotropy, as depicted in measurable quantities, such as χ (magnetic susceptibility), $M_S$ (saturation magnetization), $H_{EB}$ (Exchange Bias field) and $H_C$ (Coercive field). The dc magnetic susceptibility *χ(T)* curves (**Fig. 10d-f**), measured based on zero-field cooled (ZFC) and field cooled (FC) protocols,[70] show a broad maximum, indicative of the characteristic temperature, $T_B$, separating the superparamagnetic from the blocked state, as a result of the competition between the thermal energy and the magnetocrystalline anisotropy on the nanoscale, and is commonly expressed to be proportional to the particle volume and magnetic anisotropy.[71] Here, for similar particle volumes, $T_B$ seems to be governed by the Co content and is increasing in a nearly linear way (**Figure S7, ESI**), probably due to increased magnetocrystalline anisotropy caused by the incorporation of Co in the lattice. Additionally the ZFC part of the *χ(T)* curves presents a sudden drop in all samples (**Fig. 10 d-f)**, which resembles the paramagnetic to antiferromagnetic (AFM) transition, marked by the characteristic Néel temperature, $T_N$. The latter, progressively shifts from a low-temperature transition, as in wüstite (cf. FeO, $T_N$ = 198 K) towards room temperature, with Co-level increase (cf. CoO, $T_N$ = 291 K).[72] It is peculiar though to think whether the subtle anomaly at 125-150 K in the low-Co% sample S12, with an increased volume fraction of spinel shell, could be linked to a Verwey-like transition ($T_V$), usually observed in bulk magnetite,[73] as well as in its nanostructured counterparts.[74,75] Actually, similar Verwey-like transitions at various characteristic temperatures for Mn-, Zn-, and Co-substituted spinel ferrites have been reported.[76]

In view of assessing coupling of lattice to magnetic transitions, the temperature variation of the refined lattice parameters from xPDF is considered. For the spinel phase, they show a systematic evolution with increasing temperature for all samples (**Fig. 10 g-i**). Namely, an initial drop at ~50 K until they reach a minimum between 120 K and 150 K and then a constant expansion with further T increase. It is worth noting that the more intense, step-like drop of the lattice parameter for sample S35 at 50 K coincides with a subtle anomaly shown in the ZFC *χ(T)* around the same temperature. It is however unclear if these barely visible anomalies in the magnetization curves are indeed related to delicate magnetoelastic effects, as the resolution of our magnetometer could not permit identifying such systematics across all samples studied. However, the lattice parameters of the rock salt core present also a similar anomaly, in the same T-range as that corresponding to the spinel phase. As the two crystallographic phases are mechanically coupled, magnetostriction (cf. compressive strain) may operate through their common inorganic interface to mediate the cross-coupling between the distinct structures.

Additionally, hysteresis loops were recorded under field-cooling conditions to further support the development of a macroscopic, exchange bias field ($H_{EB}$) caused by interfacial exchange interactions. To further investigate the processes leading to $H_{EB}$, *M(H)* loops under gradually increasing cooling fields have been recorded for selected NP samples. **Fig. 11(a)** shows the evolution of $H_{EB}$ with increasing cooling fields for samples S12 and S21, compared to an earlier studied 15 nm core@shell spherical $FeO@Fe_3O_4$ sample (identified as S15 in in reference[34]). A quite peculiar observation results from their comparison**.** In S12 $H_{EB}$ values are much higher than these of S15, as expected from the better-preserved AFM core and the subsequently larger interface between the AFM and FiM phases, further assisted by the possible magnetic inhomogeneities stemming from higher magnetocrystalline anisotropy due to Co substitution. There is however, an important diminution of $H_{EB}$ for sample S21, which presents even weaker exchange-bias than the non-Co-containing iron oxide sample. Furthermore, while all samples present a weak discontinuous step like variation of magnetization near zero field (**Fig. 10 a-c**), only sample S12, compared to S21, shows two switching field distributions, marked by different maxima in *dM/dH* ZFC and FC protocols (**ESI, Figure S9**), suggesting an inhomogeneous magnetic behavior possibly arising from coexisting magnetic components of contrasting $H_c$s. Summarizing, magnetization characteristics, such as $\chi$, $M_S$, $T_B$ appear to present the expected, Co-mediated behaviour for such particles, which is mainly related to the spinel phase.[35,77,18] On the other hand, the morphological and structural study of S12 and S21 (HRTEM and xPDF) suggests that both samples have a rather narrow size distribution, similar average sizes and core to shell volume ratios (**Fig. 3**,**Table S3 in ESI**), which merely wouldn't give rise to such inhomogeneities in magnetism and the differences in the $H_{EB}$s.

One can however argue that the outcomes are a likely consequence of the clustering of MNPs that perturbs the intrinsic magnetic properties due to dipolar interactions. We are of the opinion that such an effect may be negligible, because (i) even when the "concentration" of MNPs is the largest (e.g. S35: take into account the mass of cotton wool and nanoparticle powder used, 10.3 vs. 14.6 mg), the estimated volume of the nanoparticle powder is approximately 2% of the volume of cotton wool, inferring a large average distance of the embedded in the cotton wool NCs and (ii) the emerging anisotropy contribution due to Co and vacancies (*vide infra*: Monte Carlo simulations) plays the dominant role in the magnetic behaviour of the MNPs and masks the interparticle interactions even in the lower anisotropy case. Thus, a cause for the observed differences must be sought in view of the lattice compositional variations imposed by Co-substitutions and the different distribution of defects (in terms of Fe-vacancies) in the nanoparticle volume.

**Monte Carlo Modeling of MNPs**

**Evolution of macroscopic magnetic parameters**

In order to rationalise the physical property evolution, we utilized Monte Carlo (MC) simulations to investigate how Co substitution in the MNPs' crystal lattices, influence (i) the core/shell volume fraction ratio, (ii) the atomic-scale defects (Fe/Co-lattice site vacancies) and (iii) their distribution in the nanoparticle volume, that altogether may be key parameters that couple to magnetism. MC calculations offer the exchange bias field ($H_{EB}$) and the coercive field ($H_C$) as a function of the applied cooling field strength, for low-T (**Fig. 11-b**). The parameters in the simulated models (**Fig. 1**) were chosen in such a way that we could investigate whether the experimentally observed trend, depicted in **Fig. 11-a** is reproduced when Co substitution level increases. The core/shell volume ratio, the relative amount of vacancies and their phase preference (core or shell) in these models, are considered as Co-related effects and constitute a simplified, representative illustration of the experimental observations and the conclusions made from the xPDF analysis.

The corresponding amount and type of spins in each region of the three simulated MNPs ("pure", model#1, and model#2; **Fig. 1**) are given in table **S1 in ESI**. $H_{EB}$ is directly related to the applied $H_{\mathrm{cool}}$, as a result of uncompensated spins frozen at the core-shell interface. There is a sharp increase in the $H_{EB}$, when starting from zero-field and gradually applying larger $H_{\mathrm{cool}}$ (**Fig. 11**). This is attributed to the competition between the Zeeman energy and the other magnetic interactions within the system. As the $H_{\mathrm{cool}}$ increases, the magnetic coupling between the field and the magnetic moments (Zeeman energy) increases, tending to orient them along the field direction. For high enough fields, such coupling competes with the other magnetic interactions within the system overcoming the exchange coupling at the interface, which leads to a peak and then a gradual decrease of the $H_{EB}$ when the Zeeman energy dominates. This decrease is mainly determined by the intrinsic magnetic anisotropy of the system and by its microstructure.[53],[78] Differences between the observed and the MC calculated results are attributed to the fact that the model takes into account only the essential characteristics of the particle microstructure. $H_C$ on the other hand, although similarly affected by Zeeman energy, is much less dependent on the $H_{cool}$ because it is governed largely by the magnetocrystalline anisotropy in both magnetic phases, thus allowing for relatively high values even at $H_{Cool} = 0$. Experimental and MC calculated results presented in **Fig. 11** are in line with the described mechanisms. For an in-depth understanding of the mechanisms that dominate these measurable quantities, one has to look more closely into the details of crystal lattice configuration of the simulated models. (**Fig. 1, Section S1 in ESI**)

Apart from the competition of the spins at the core-shell (AFM-FiM) interface, some spins in the AFM phase (closed to-, but not necessarily at the interface) can contribute to the exchange bias. Such spins are pinned to a certain direction after the field cooling, and are coupled to the FiM layer.[79] The origin of these pinned moments is not yet quite clear, but they are believed to be related with some kind of disorder, such as point defects.[80] In fact, weak $H_{EB}$s have been measured in the past for highly defected (due to crystallographic vacancies), single phase sub-critical size (d~ 10 nm) FiM NCs.[34] In order to give an estimate of these effects in the MC calculation (**Fig. 11 b**), the contribution of defects accounted in these models was also examined (**Fig. 1**,**Table S1 in ESI**). So, upon Co-substitution:

**model#1**, represents particles with a significant amount of defects in their AFM core and core-IF, acting as pinning centres. The shell, with no defects at all, adopts a perfect FiM ordering. Additionally, the shell's large volume fraction in relation to the total particle volume, means that the pinned moments around the defects should exhibit a strong coupling with the easily magnetized FiM phase, resulting in a large exchange bias. The inhomogenous reversal of spins found in such exchange-bias systems leads to an increased $H_C$ as well,[81] illustrated also in **Fig. 11 b** for model#1.

"**pure**" model, offers a somewhat lower exchange bias and coercive fields. Although this model has a similar AFM volume (20% core compared to 25% core in model#1), the significantly smaller population of defects in the core and the core-IF results in less pinning centres and consequently exchange interactions, lowering the $H_{EB}$ and $H_C$. A more defected FiM shell should also negatively affect its ability to couple with pinned spins near the AFM/FiM interface.

**model#2**, presents the lowest calculated $H_{EB}$. In this case, there are no defects at all in the AFM core, which could play the role of pinning centres. The increased volume fraction of core (50%) results in a relatively thin and highly defected FiM shell, which along with a high Co percentage (Co substitution decreases the magnetization of spinel: **Fig. 10 d-f** , **Table S4 in ESI**) is unable to strongly interact with the uncompensated spins at the AFM/FiM interface. Thus, such a small $H_{EB}$, as shown in **Fig. 11 b**, is the outcome. Normally, an increased Co-level in a given Fe/Co spinel system would increase its coercivity, due to the increased magnetocrystalline anisotropy, mediated by Co. This is not the case in mode#2 because of the discussed morphological characteristics (smaller volume of spinel-FiM phase, which has a rather negligible contribution in $H_C$) and the very weak exchange coupling interactions.

**Towards optimally performing MNPs**

The discussed behaviour of the simulated models is in good agreement with the experimental trends. In that respect, sample S12 presents the highest $H_{EB}$ and $H_C$ values among all samples, since it has a rather defected core, surrounded by a not so defective spinel shell, involving a relatively low Co-content, as shown by the structural analysis. On the other hand, sample S21 shows a weaker performance, in terms of $H_{EB}$ and $H_C$, which is explained by its less defected and larger core, surrounded by a more defective shell together with a higher Co-substitution rate. The S15, would be expected to perform in a similar manner to S12. It turns out though, that since heterostructured $FeO@Fe_3O_4$ NCs[40] are not able to preserve the oxidation-sensitive core[34] (compared to Fe/Co-bearing MNPs, with similar structural and morphological characteristics, e.g. S12), the AFM/FiM interface is compromised as well as the ability of S15 for larger exchange coupling.

To shed light on the effects of Co-substitution and defects (vacant sites), on $H_{EB}$ and $H_C$, we further investigated several variations of model #1 and model #2, looking into how core or shell separately may be affected (**S1, Figure S1, Figure S2 in ESI**). A particle with a perfectly ordered, non-defective core would exhibit the poorest magnetic response. However, this extreme case is rather impossible to obtain in a lab-based chemical synthesis protocol. Realistically, the rock salt core plays mainly a role in the exchange interactions to the extent that it creates a well-defined AFM/FiM interface, so, it is crucial that it can be sufficiently protected from oxidation and preserved over time. On the other hand, the spinel shell seems to be strongly, negatively impacted by a large population of defects, because these perturb the FiM ordering, and thus the shell's ability to interact with uncompensated or pinned spins in the core, near the interface. We find that raising Co-concentration has favourable effect on the properties of interest (**Figure S2, ESI**), including counter-balancing the detrimental defect-induced effects.

Therefore, designing plausible optimized core@shell MNPs, where sizeable $H_{EB}$ and $H_C$ are combined, while high magnetization values are maintained, would require devising NCs, bearing key characteristics, namely: (i) grown in the size range close to 15 nm, (ii) have a moderate Co substitution level (~10%), (iii) involve a well-preserved core, accounting for a low-fraction (~20-30%) of the entire particle volume, with defects typical of a sub-stoichiometric rock-salt phase, and (iv) a spinel shell that is as little defected as possible, preferably resembling the ideal spinel structure. It is not surprising that S12, which presents a desired magnetic behavior (**Fig. 10 d** and**Fig. 11 a**), has such characteristics, making this sample a nearly optimal candidate for applications.

**Calculation of the Specific Absorption Rate (SAR)**

The measured magnetic properties and the associated lattice-driven effects, discussed in the previous sections, aimed to set the stage for assessing the MNPs' application potential as smart heating mediators. The possibility for exploiting magnetic heating, generated by ferrite particles exposed to AC magnetic fields, offers an innovative modality in biomedical applications. To this extent magnetic hyperthermia treatment of tumours, has been extensively studied in recent years.[82] Ferrite particles have been widely considered for such applications, as their heating power shows a marked dependency on particle size and high sensitivity in size-distribution met in real samples.[11] In that respect, the largest among them (d~ 25 nm), display enhanced magnetically-induced heating performances, especially when structural, defects occurring during synthesis are eliminated.[30] However, under particular circumstances, during the oxidative transformation of $Fe_xO@Fe_3O_4$ core-shell nanocubes (d~ 23nm) atomic scale defects, such as $Fe^{2+}$ deficient sites, may positively influence hyperthermic properties.[83] Along this direction, raising the population of Fe-vacancies, in small size $Fe_xO@Fe_3O_4$ core-shell nanospheres (d~ 8 nm), seem to offer an alternative exploitable pathway for magnetic hyperthermia.[34] The differences in the observed heating responses are generally related to critical particle sizes[84] and magnetic anisotropy terms.[85] Thus, optimizing the latter for particle sizes, tailored within the biological compatible limits, set by toxicity and patient discomfort,[86] is particularly important. In this endeavor, heat generation capabilities can be assisted further by tuning the MNPs' morphology,[55] include topologically arranged coupled core and shell phases,[12] increase cation deficiency or even control composition.

The detailed characterization of the $Co_yFe_{1-y}O@Co_xFe_{3-x}O_4$ IONCs, involving properties as $H_{EB}$ and $H_C$, pertaining to exchange and magnetocrystalline anisotropies, aimed at exploring possible complementary pathways for enhancing the heating performance of exchange coupled AFM/FiM phases. For this purpose, we calculated the SAR values for the studied MNPs, according to the Linear Response Theory for the Néel-Brown relaxation model.[10] (**S2, ESI**) The calculations were performed for different AC field amplitudes, $H_0$, at a frequency of $f$ = 500 kHz. The SAR performance for MNPs with various structural characteristics is compiled in **Fig. 12.** The calculations compare Co-substituted (0-35%) mixed-metal oxides, developed for this work, and two contrasting iron oxide particles, with no Co-incorporation, reported earlier. The latter are either of a similar to the Co-derivatives size, with $Fe_xO@Fe_3O_4$ core@shell (S15; d~ 15 nm) structure, or of a sub-critical size, involving single-phase (S8; d~ 8 nm), highly-defected ferrospinel NCs.[34]

Since SAR shows a marked dependency on particle size, the size distribution in real samples is crucial when comparisons are drawn. For example, experiments on cobalt ferrite MNPs (8-25 nm) find SAR values ranging from 40-50 W/g in small (8 nm) and depending on the conditions ($H_0$, $f$), raise up to about 600-800 W/g at intermediate sizes (13-15 nm).[87] To place the present work in the context of related experiments, we would like to note that our calculated magnetic fluid heating power of 300-400 W/g (**Fig. 12**) for the present core@shell nanoparticles, are for example, within the window spread of measured SAR values, obtained from polydisperse iron-oxide (cf. maghemite and cobalt ferrite) nanoparticles,[11] with characteristic diameters between 12-17 nm (i.e. similar to our MNPs), studied also under related conditions ($H_0$= 24.8 kA/m and $f$= 700 kHz).

The calculated SAR values follow a smooth monotonic increase (**Fig. 12**), resembling the experimentally measured upward trends reported previously.[84] The higher anisotropy volume of MNPs with core@shell topology results in a deviation from the quadratic field dependence of the SAR, [87] displayed only by the smaller particle (S8). In one hand, increasing the anisotropy of ferrite MNPs, consisting of a single magnetic phase (S8) by introducing a large amount of defects and thus spin-pining centers, makes such particles to perform comparably to the $Fe_xO@Fe_3O_4$ system (S15) carrying higher exchange anisotropy (cf. S8-S15,**Fig. 12**). However, when the magnetocrystalline anisotropy is raised by introducing Co ions in the lattice, in addition to an already strongly exchange-coupled system, the resulting SAR performance is assisted further (cf. S12,**Fig. 12**), thus supporting our aforementioned hypothesis for a nearly optimally designed candidate (S12). Worth noting though, that a simple coexistence of Co ions and AFM/FiM interfaces does not warrant enhanced SAR. Our calculations indicate that high magnetocrystalline anisotropies, usually expressed by large $H_C$s, and high exchange anisotropies, supported by sizeable $H_{EB}$s, are prerequisites for optimal SAR values. In fact, the evolution of the calculated SARs (cf. S12> S15> S21; **Fig. 12**) follows the trend obeyed by the related magnetic quantities (**Fig. 11**) that are mediated by compositional changes (i.e. Co-content and vacancies). Further Co-content increase, generates an even weaker SAR performance (S35), due to the Co-mediated nanocrystal transformations and the consequent $H_{EB}$ and $H_C$ weakening, already discussed.

**Conclusions**

The effects of cobalt incorporation in spherical core@shell $Co_yFe_{1-y}O@Co_xFe_{3-x}O_4$ NCs (d~ 15 nm) have been studied on atomistic level. We find that the sub-stoichiometric rock salt-like Wüstite ($Fe_yO$) tends to heal its vacant Fe sites upon Co substitution, helping its stabilization. The trend becomes more prominent at elevated Co levels that appear to impede the mechanism for oxidative conversion of rock salt to spinel. Self-passivation of the initially formed rock salt NCs is still possible, creating a spinel-like shell, which yields core@shell nanostructures with relatively high volumetric ratios of the core phase that grows with the Co content. The xPDF analysis indicate that a higher divalent metal ($M^{2+}$) stoichiometry in the rock salt part by Co-substitutions, results in increased levels of unoccupied tetrahedral (Td) sites in the spinel, likely through oxidative shell creation involving an ordered defect-clustering mechanism. The population of Td site atomic-scale defects of the shell is directly correlated to the core stabilization and increases when Co abundance increases. On the other hand, the extent of octahedral (Oh) vacant sites seems to follow the more conventional behavior met in ferrospinel nanostructured systems.

Concerning the magnetic properties, the core-to-shell volumetric ratio is shown to have an immediate effect on the buried interface dimensions, thus affecting the exchange coupling interactions, reflected in a sizeable exchange-bias field ($H_{EB}$). The latter shows a strong correlation with the cooling field strength and depends on the ability of the available uncompensated spins at the AFM-FiM interfaces. On the other hand, the coexistence of Co sites and atomic-scale defects in the crystal structure, in the form of Fe- and/or Co-ion Td vacancies in both magnetic phases, significantly affects the magnetocrystalline anisotropy and magnetic ordering. We find that raising Co-concentration has favorable effect on the properties of interest, including changes in the magnetization values and coercive fields ($H_C$). In fact, the enhanced $H_C$ and $H_{EB}$ seem to be prerequisites for hyperthermic effects entailing high Specific Absorption Rates (SAR).

Combining experiments with Monte Carlo simulations, we are able to suggest that designed heterostructured NCs of about 15 nm total diameter (cf. below the critical size, d< 20-30 nm), synthesized by controlled colloidal chemistry protocols, with Co content level of about 10%, which leads to a volumetric fraction of the core to about 20-30% (over the whole particle volume), should result in optimal heating power. To achieve such a performance, the nanoarchitercure is dressed with the advantageous protection of the spinel shell, in contrast to not that well-performing, in terms of SAR, non-Co-containing iron oxide NCs. The suggested Co substitution level allows for a certain amount of desirable defects in the rock salt core, but would not create excessive vacancies in the Td sites in the spinel, which would perturb its otherwise beneficial FiM ordering and the shell's ability to interact with uncompensated spins near the buried interface. Fine tuning of such quantities, could result in MNPs with remarkable magnetically-mediated heating power for possible hyperthermia applications.

**Author Contributions**

Investigation: GAnt – Synthesis/XRD/ TEM/ EDS; Investigation: EB, MA, GAnt – PDF; Investigation: VI & GAus – Magnetic measurements; Formal analysis: GAnt – Experiment; Formal analysis: MV & KNT – Theory; Writing – original draft: GAnt; Writing – Review & Editing: GAnt, EB, AL; Supervision: AL; Project administration: AL; Conceptualisation: AL.

**Conflicts of interest**

There are no conflicts to declare.

**Acknowledgements**

This research used the beamline 28-ID-2 of the National Synchrotron Light Source II, a U.S. Department of Energy (DOE) User Facility operated by Brookhaven National Laboratory (BNL). Work in the Condensed Matter Physics and Materials Science Division at BNL was supported by the DOE Office of Basic Energy Sciences. Both activities were supported by the DOE Office of Science under Contract No. DE-SC0012704.

**Notes and references**

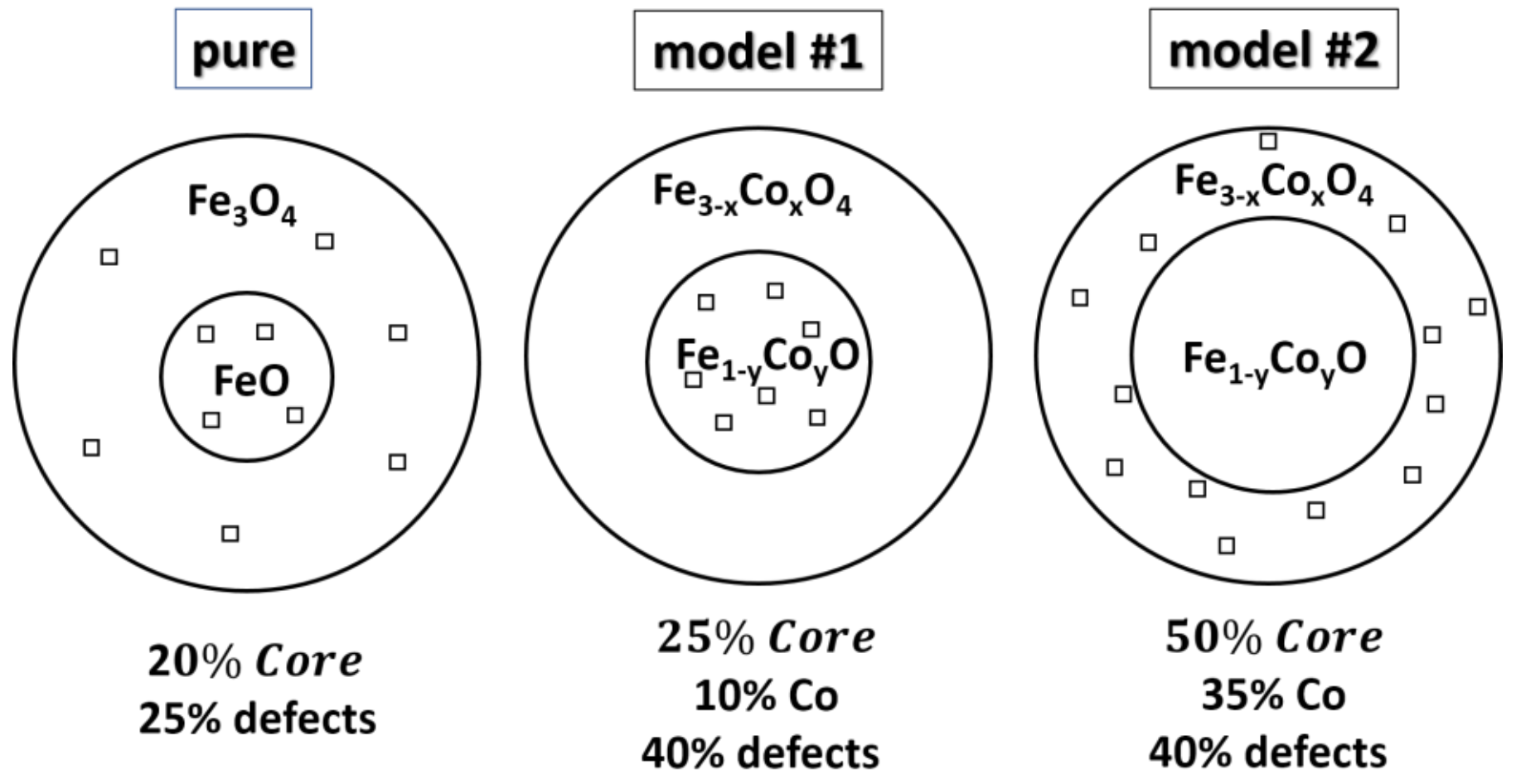

Fig. 1 Schematic, qualitative illustration of various heterostructured core@shell nanocrystals. The magnitude of the chosen parameters for three representative models, studied by Monte Carlo simulations is given, namely: relative volume fractions of core (compared to shell), Co substitution level (in percentage), population of defects in the sense of vacant metal-ion crystallographic sites (depicted with open squares) and their distribution in the nanocrystals' volume.

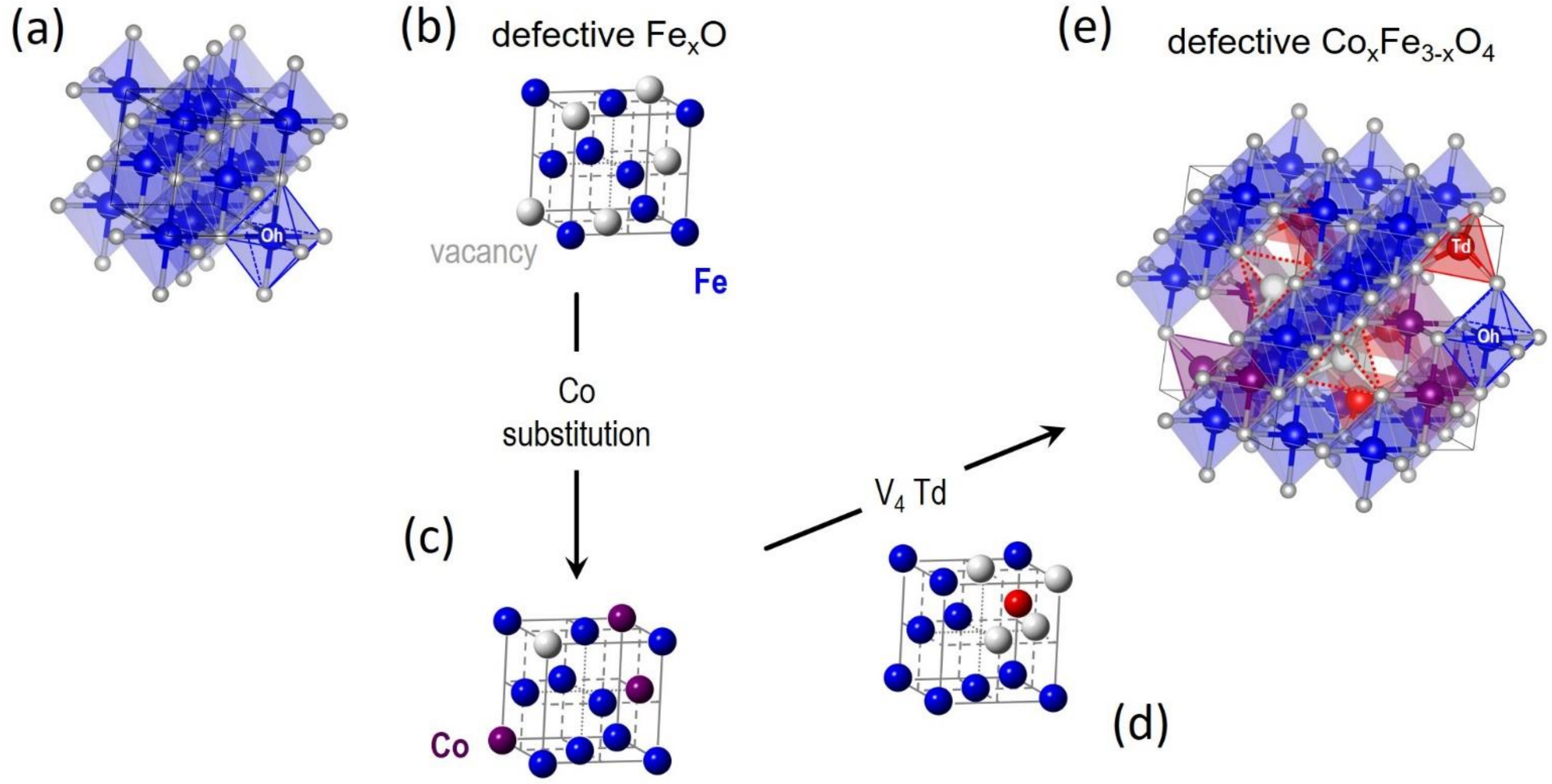

Fig. 2 Crystal structure of (a) stoichiometric rock salt wüstite (FeO), with octahedrally (Oh) coordinated by oxygen (small white spheres) ferrous $Fe^{2+}$ sites (blue spheres). The same structure, simplified not to show oxygen and coordination polyhedra, where (b) is a defected $Fe_xO$ structure, showing Fe and vacant (*V*) metal-ion sites only (large grey spheres), and (c) is a Co-(purple spheres)-substituted derivative, depicting a less-defected rock-salt structure. (d) A plausible ordered-defect rock-salt structure, with $Fe^{2+}$ Oh sites, and interstitial ferric ($Fe^{3+}$) tetrahedral sites (Td; red spheres) surrounded by four vacant (*V*) iron positions for charge balance. (e) Coalescence of $V_4$-Td defect clusters[59] upon further oxidation, may offer a likely pathway towards the nucleation of a defective (Td, vacant sites – depicted by dotted red line trigonal pyramids), Co-substituted ferrospinel structure.

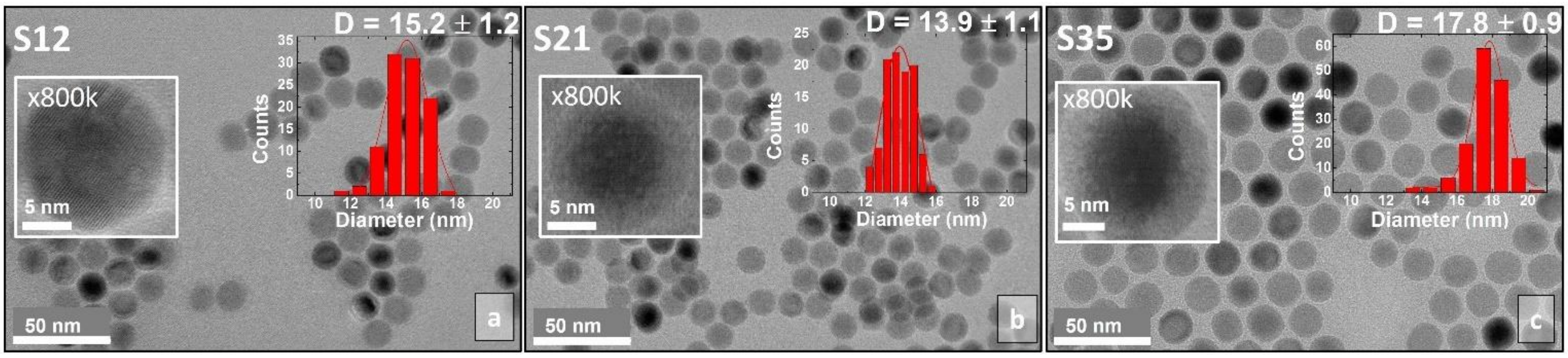

Fig. 3 Low-magnification bright-field transmission electron microscopy (TEM) images of core@shell $Co_xFe_{1-x}O@Co_yFe_{3-y}O_4$ spherical nanocrystals and insets of high-resolution TEM images of selected particles at 800k magnification. The nanoparticle size distributions are also shown as insets.

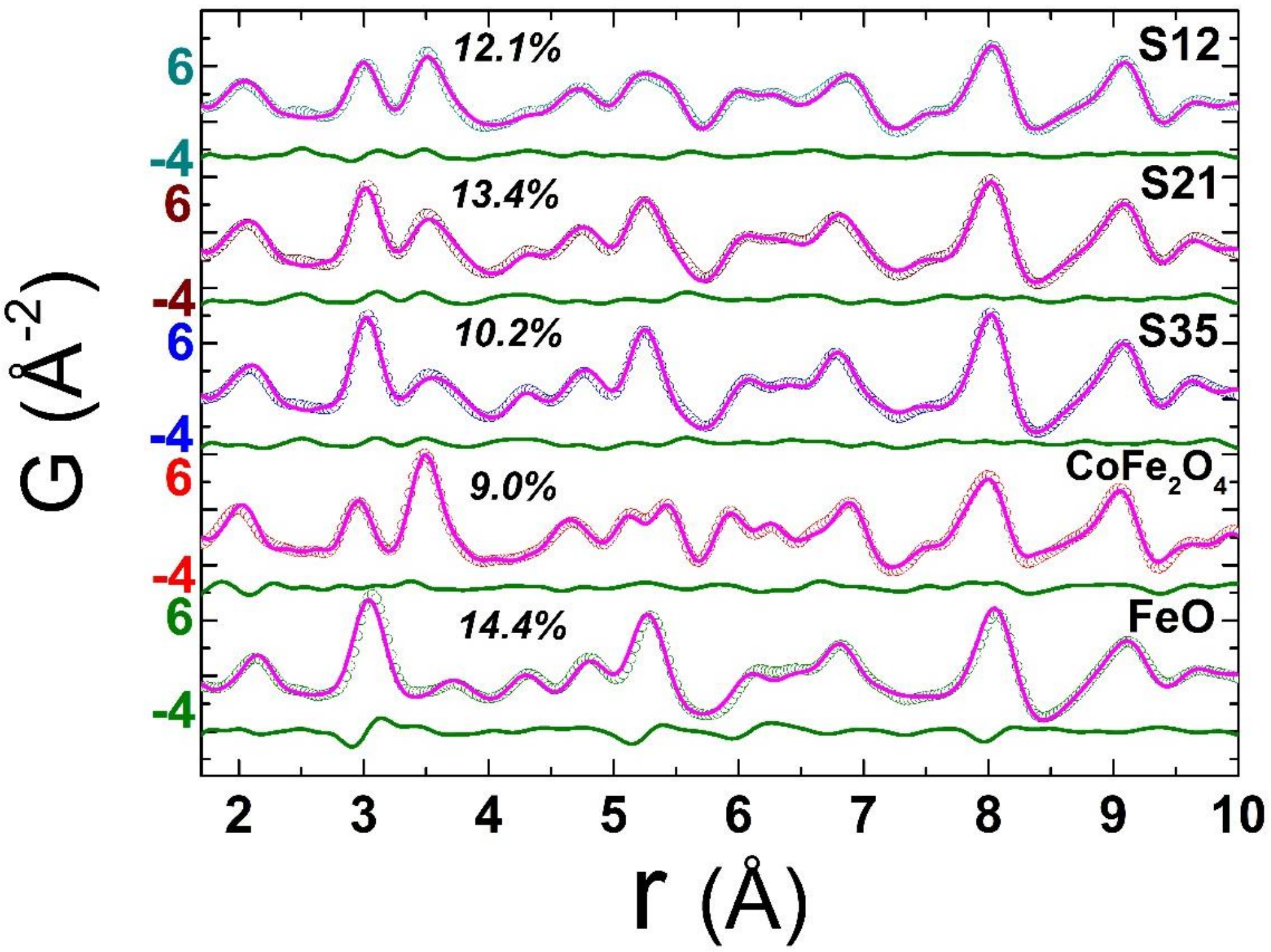

Fig. 4 Experimental atomic xPDF data at 300 K for the spherical nanocrystal samples S12, S21, S35 and the bulk reference materials of $CoFe_2O_4$ and FeO, plotted as a function of the radial distance, *r*, fitted in the low- *r* range (1.7-10 Å). The solid pink line over the data (open circles) is the best fit based on either the rock-salt FeO (Fm-3m symmetry) or the cubic-spinel $Fe_3O_4$ (Fd-3m symmetry) model for the references, respectively, or a combination of these in a two-phase model for the NCs. The quality of fit factors ($R_w$) are also given. The green line below each fit corresponds to the difference between observed and calculated PDFs. The larger $R_w$ factor (14.4%) for the fit of FeO reference, in comparison to the rest, is a likely result of its highly defected, sub-stoichiometric nature ($Fe_{1-x}O$), even in the bulk form.

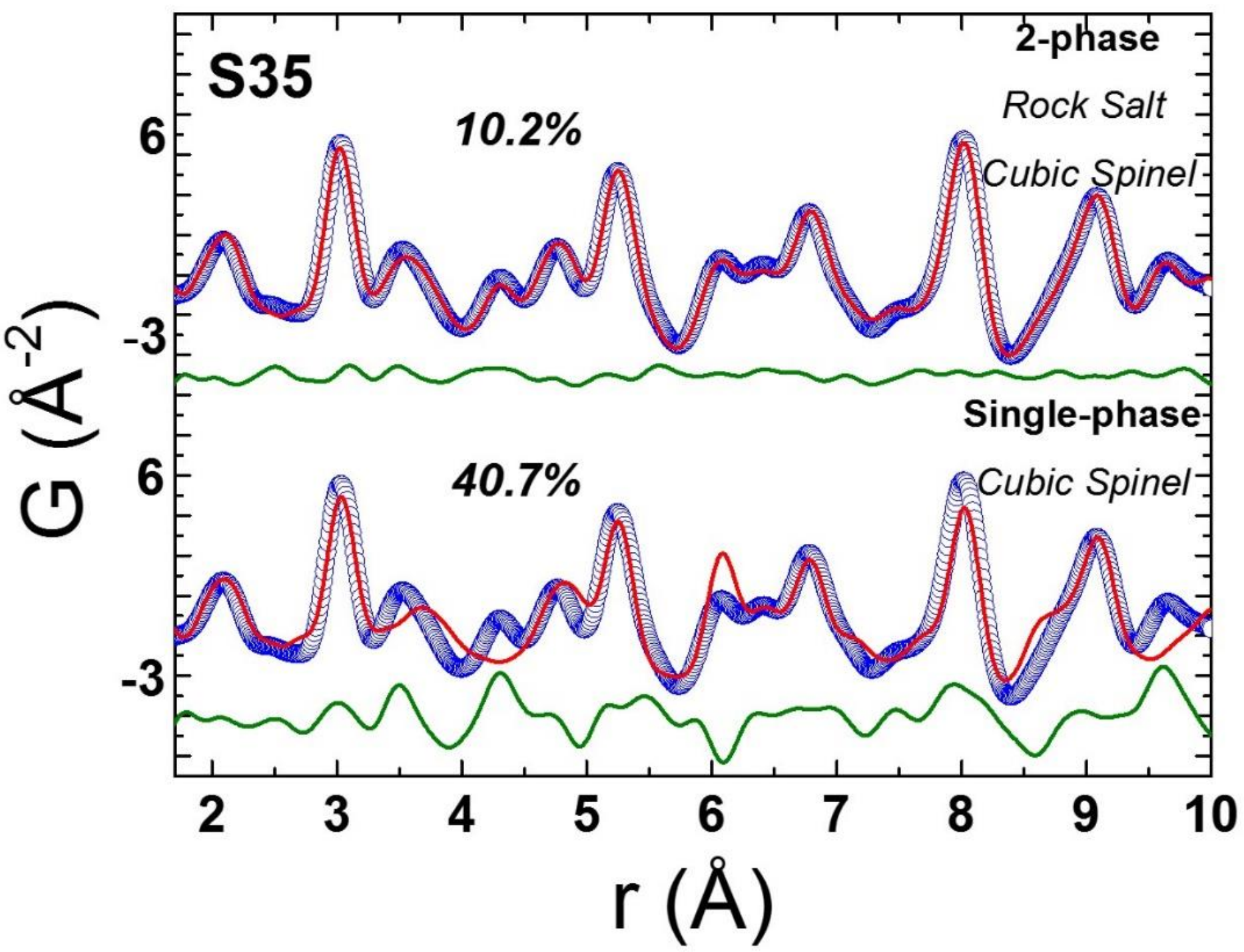

Fig. 5 Representative xPDF fits of data at T= 300 K over the low-r PDF region (1 nm) for nanocrystal sample S35, assuming a single-phase cubic spinel model (Fd-3m, $R_w$= 40.7%) at the bottom and a 2-phase rock-salt/ cubic spinel model (Fm-3m/ Fd-3m, $R_w$= 10.2%) at the top, indicating clearly a structure consisting of two crystallographic phases for nanocrystals (S35). Similar results verify the core@shell structure for the rest of NCs studied (ESI, Figure S6).

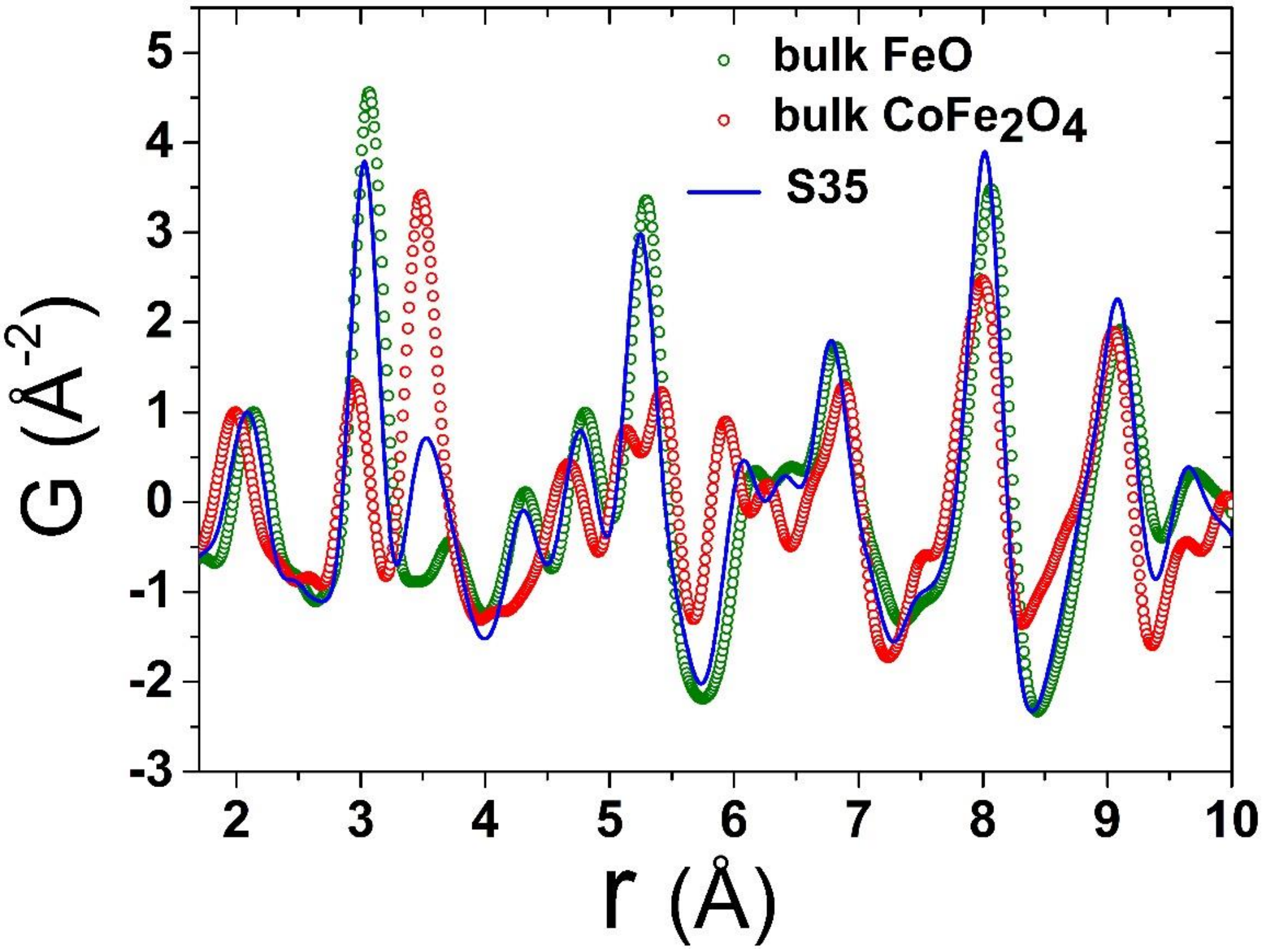

Fig. 6 Representative low-r raw data collected at T= 300 K for reference materials FeO (green circles), $CoFe_2O_4$ (red circles) and the FeO@$CoFe_2O_4$ nanocrystal sample S35 (blue line). The normalization is realized by dividing the x-PDF patterns with the intensity of the peak centered at ~2.0 Å (first peak, attributed to nearest-neighbour Fe-O pairs). The various contributions of the two different crystallographic phases, represented here by the bulk reference samples, account for slight peak shifts and, in some cases, variations of peak intensities of the experimental PDFs in the NCs. For clarity, only data for nanocrystal sample S35 are shown here, whereas data for the remaining NCs (S12, S21) are compared in Fig. 7.

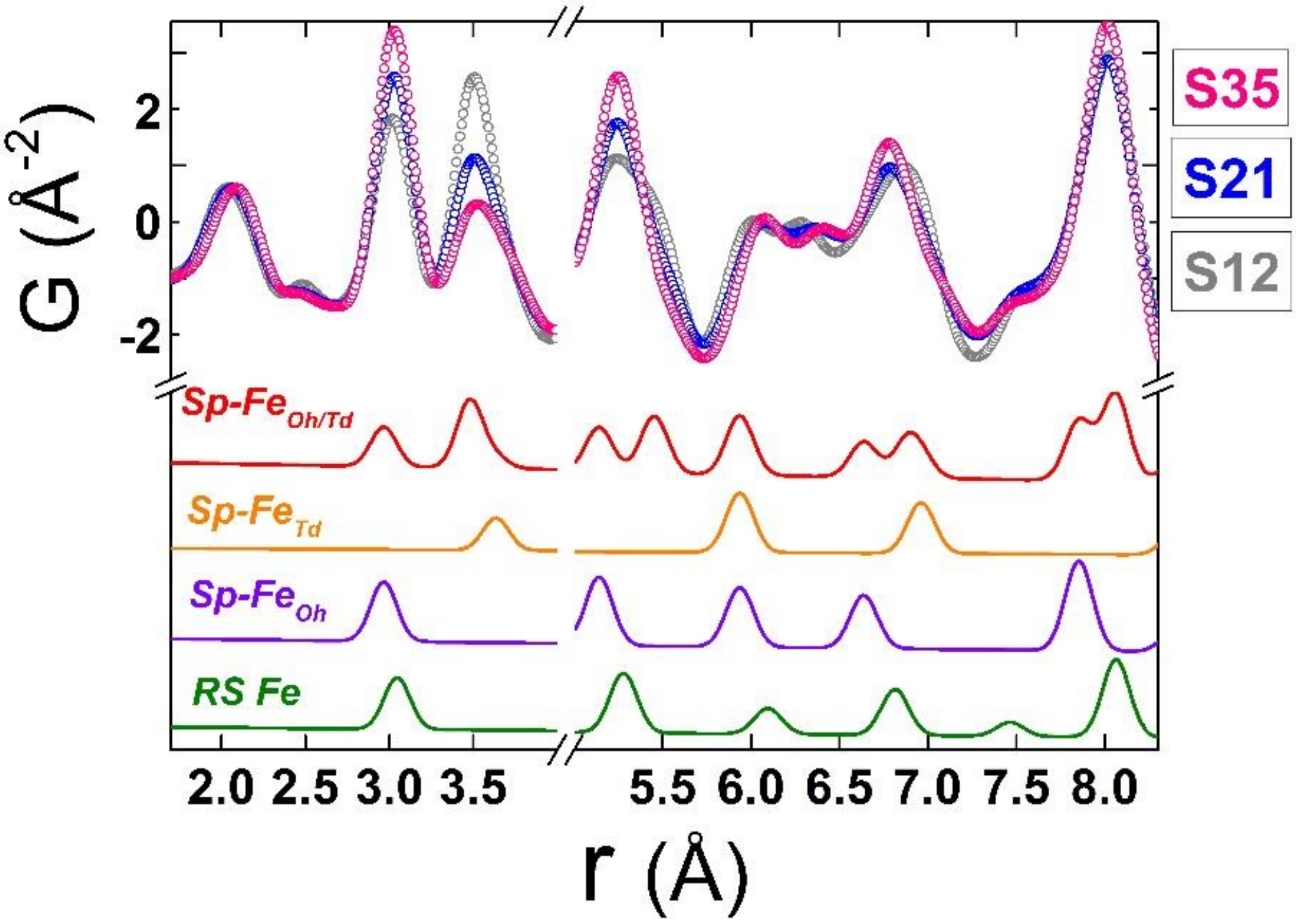

Fig. 7 Normalized low-r PDF data obtained at T=300K for all spherical nanocrystal samples, shown with coloured, open circles. The data are plotted over each other to visually enhance the differences in peak intensities, arising from different Rock-Salt *vs.* Spinel volume ratios and possible stoichiometry-related effects. The colored solid lines beneath, represent the calculated partial PDF contributions arising from interatomic distances of Fe-Fe pairs only in a rock-salt cell, containing only Fe atoms (RS Fe: green line), in a spinel cell containing only octahedral Fe atoms (Sp-$Fe_{Oh}$: purple line), a spinel cell containing only tetrahedral Fe atoms (Sp-$Fe_{Td}$: yellow line) and a spinel cell containing both octahedral and tetrahedral atoms (Sp-$Fe_{Oh/Td}$: red line). Calculations are based on the aforementioned models for rock-salt and normal cubic spinel configurations; the oxygen sub-lattices are neglected for simplicity. The obtained simulated profiles are arbitrarily scaled in the graph, for a better visualization of the expected positions of Fe-Fe bond distances in the relevant phases.

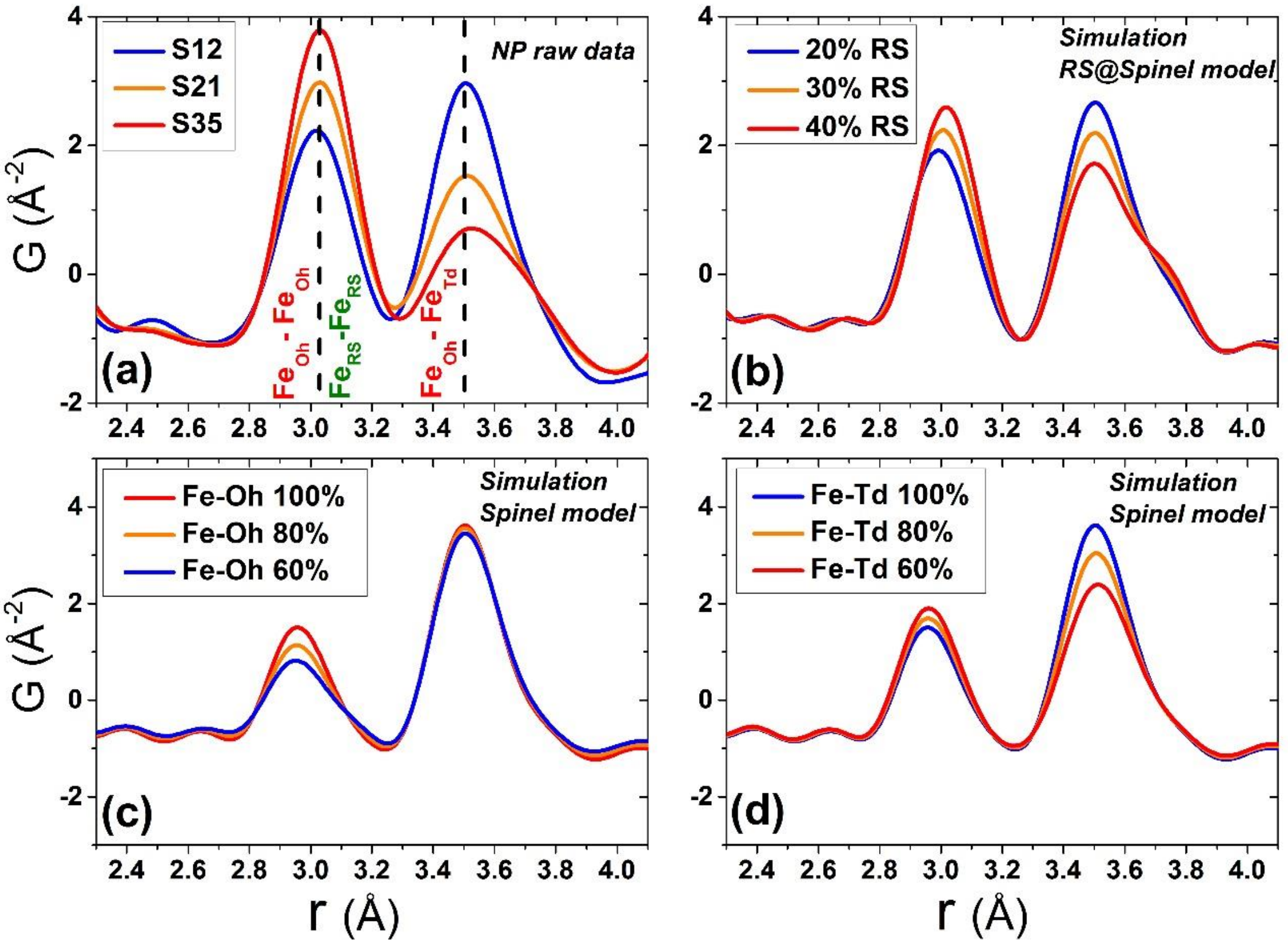

Fig. 8 (a) Normalized experimental G(r) patterns at T= 300 K in the < 5 Å r-region for all nanocrystal samples, focusing on the peaks corresponding to $Fe_{Oh}$-$Fe_{Oh}$ (~3 Å) and $Fe_{Oh}$-$Fe_{Td}$ bond distances (~3.5 Å) in the spinel crystal lattice and Fe-Fe bond distances (~3 Å) in the rock-salt lattice. Simulated xPDF patterns on the basis of: (b) the 2-phase Rock-salt/cubic Spinel model, calculated for different rock-salt to spinel phase volume ratio; (c) a single-phase cubic spinel model, assuming fully occupied tetrahedraly (Td) coordinated Fe atoms, while the octahedral (Oh) Fe-site occupancy is varied in a stepwise manner, starting from 60% up to 100% (fully occupied); (d) a single-phase cubic spinel model, assuming fully occupied Oh Fe-sites, while varying the Td Fe-site occupancy up to 100%. The effects of the various configurations used in the simulations, reflect on the relative intensity ratio modifications of the PDF peaks, at ~3 Å and ~3.5 Å.

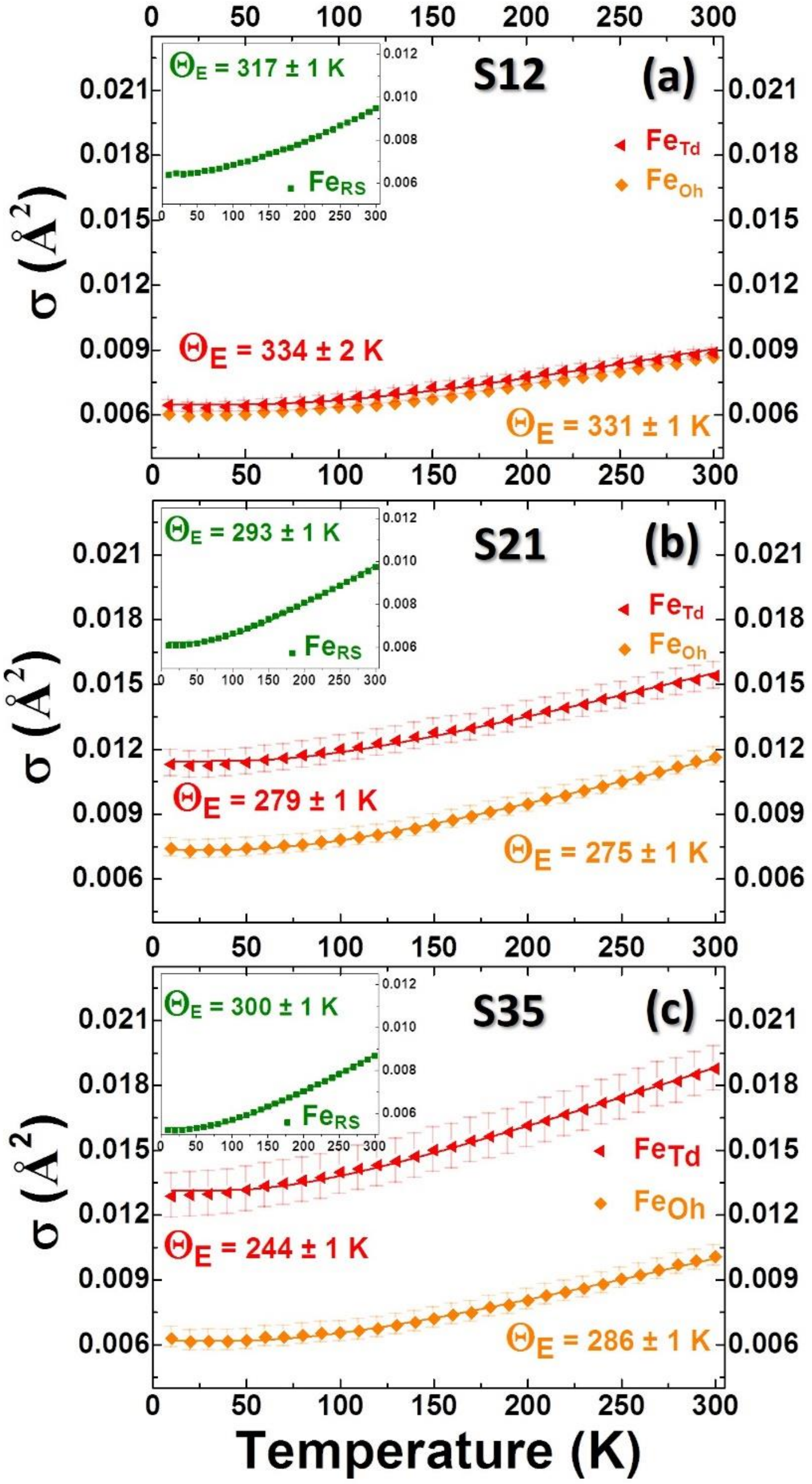

Fig. 9 Refined isotropic temperature parameters (σ) of Fe atoms in rock-salt (RS) (Insets: green points), $Fe_{Oh}$ (orange points) and $Fe_{Td}$ (red points) in spinel, derived from refinements of experimental PDFs in the 10-300 K temperature range, for nanocrystal samples S12, S21 and S35, on the basis of a typical 2-phase model of a rock-salt/spinel configuration. Error bars are shown, although for RS they are smaller than the actual symbols. The T-dependent temperature parameters were further analysed in view of the correlated Einstein model (solid lines) to obtain the Einstein Temperature, $\Theta_E$ and evaluate the average static thermal displacements, $\sigma_0$ (see text for details).

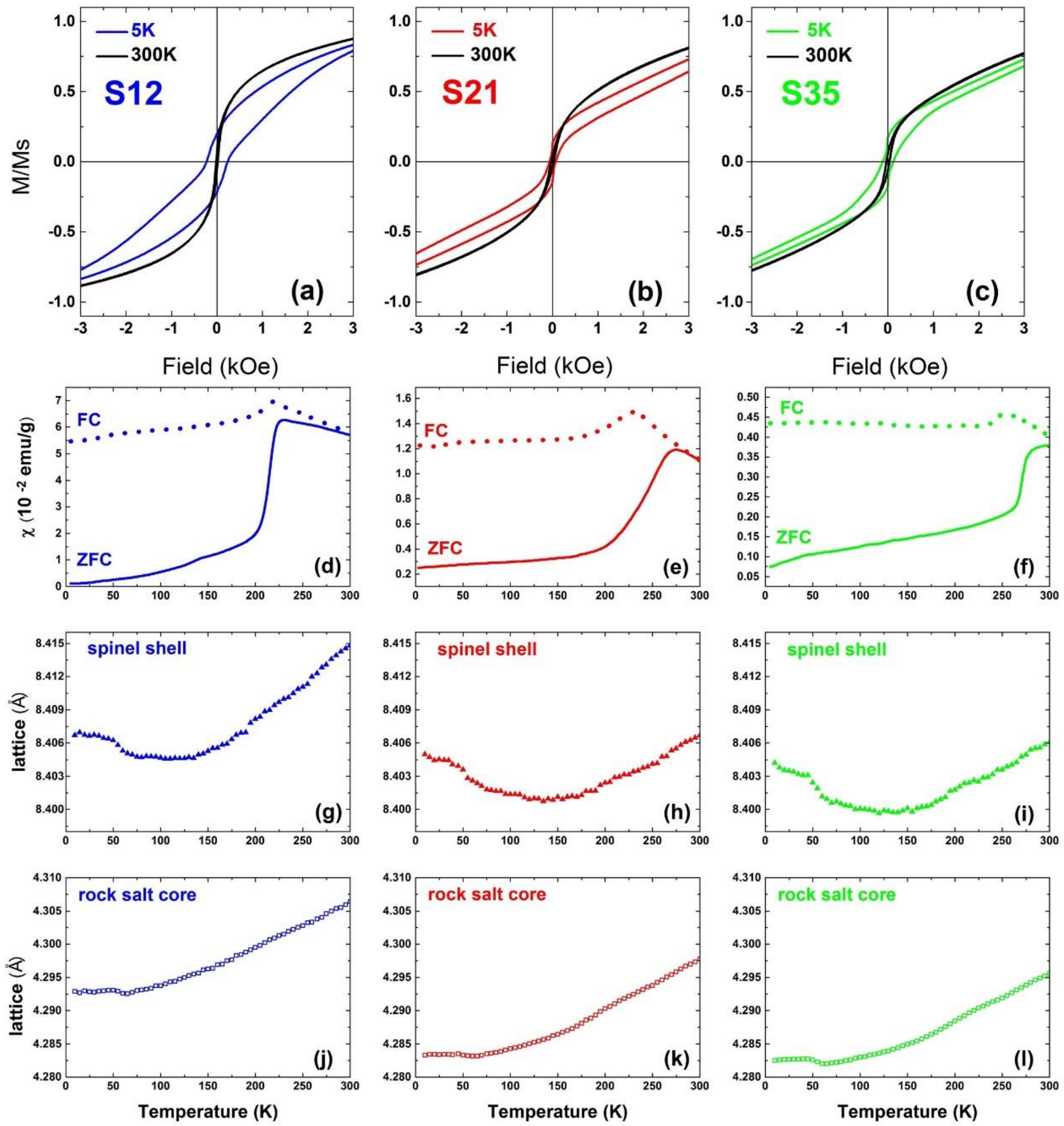

Fig. 10 The low-field part of the normalized hysteresis loops ($M/M_S$) at 5 K and 300 K for nanocrystal samples S12 (a), S21 (b) and S35 (c) taken under zero field-cooled protocols. Temperature evolution of the zero-field cooled (ZFC, solid lines) and field-cooled (FC, dotted lines) susceptibility curves for the nanocrystals S12 (d), S21 (e) and S35 (f), under a magnetic field of 50 Oe. Lattice parameter temperature evolution in the same T-range for core (j-l) and shell (g-i) crystallographic phases.

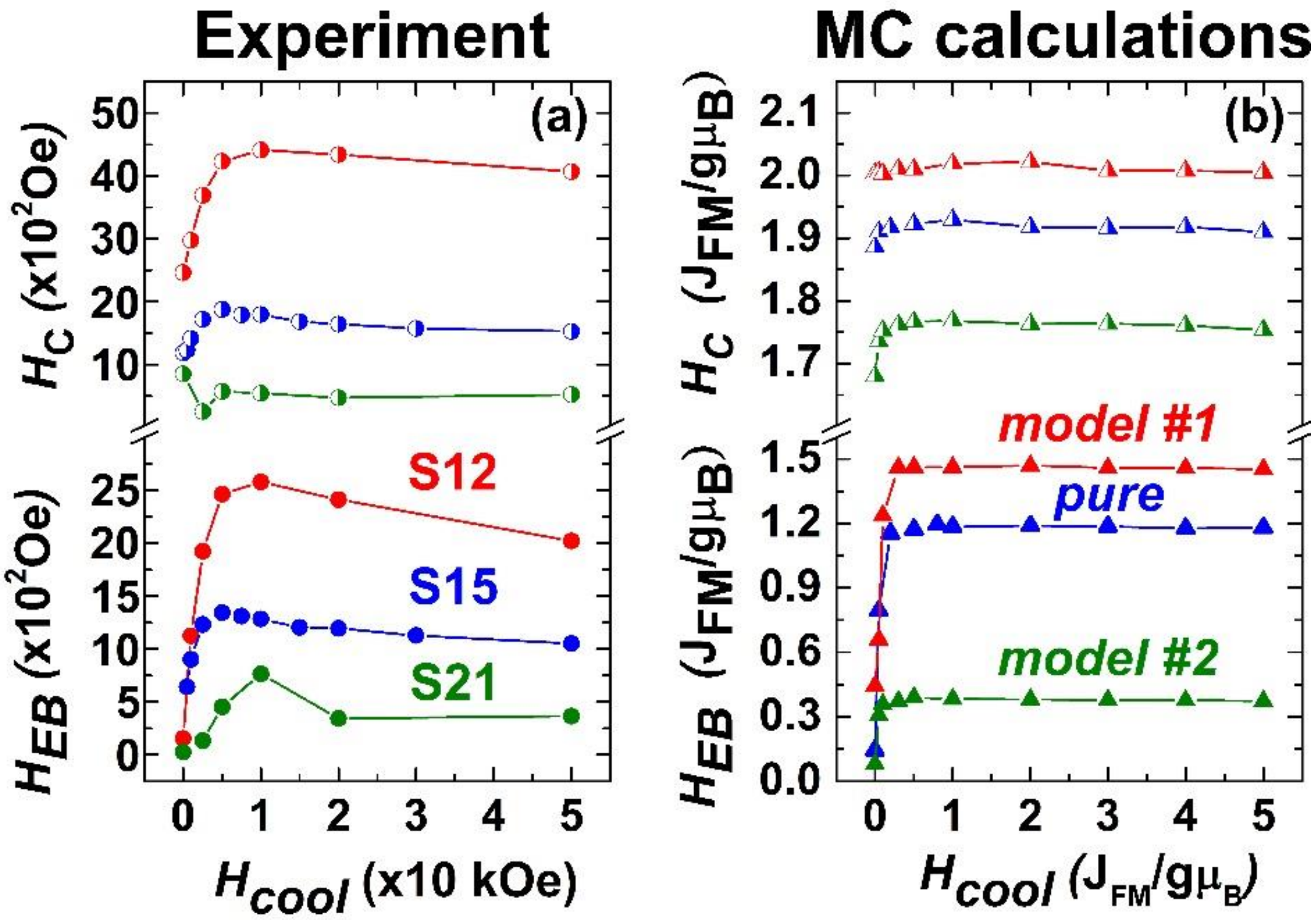

Fig. 11 (a) Experimentally determined exchange bias ($H_{EB}$, *full circles*) and coercive fields ($H_C$, *half-full circles*) and (b) Monte Carlo calculations of $H_{EB}$ (full triangles) and $H_C$ (half-full triangles) for the three heterostructured nanocrystal models (Fig. 1).

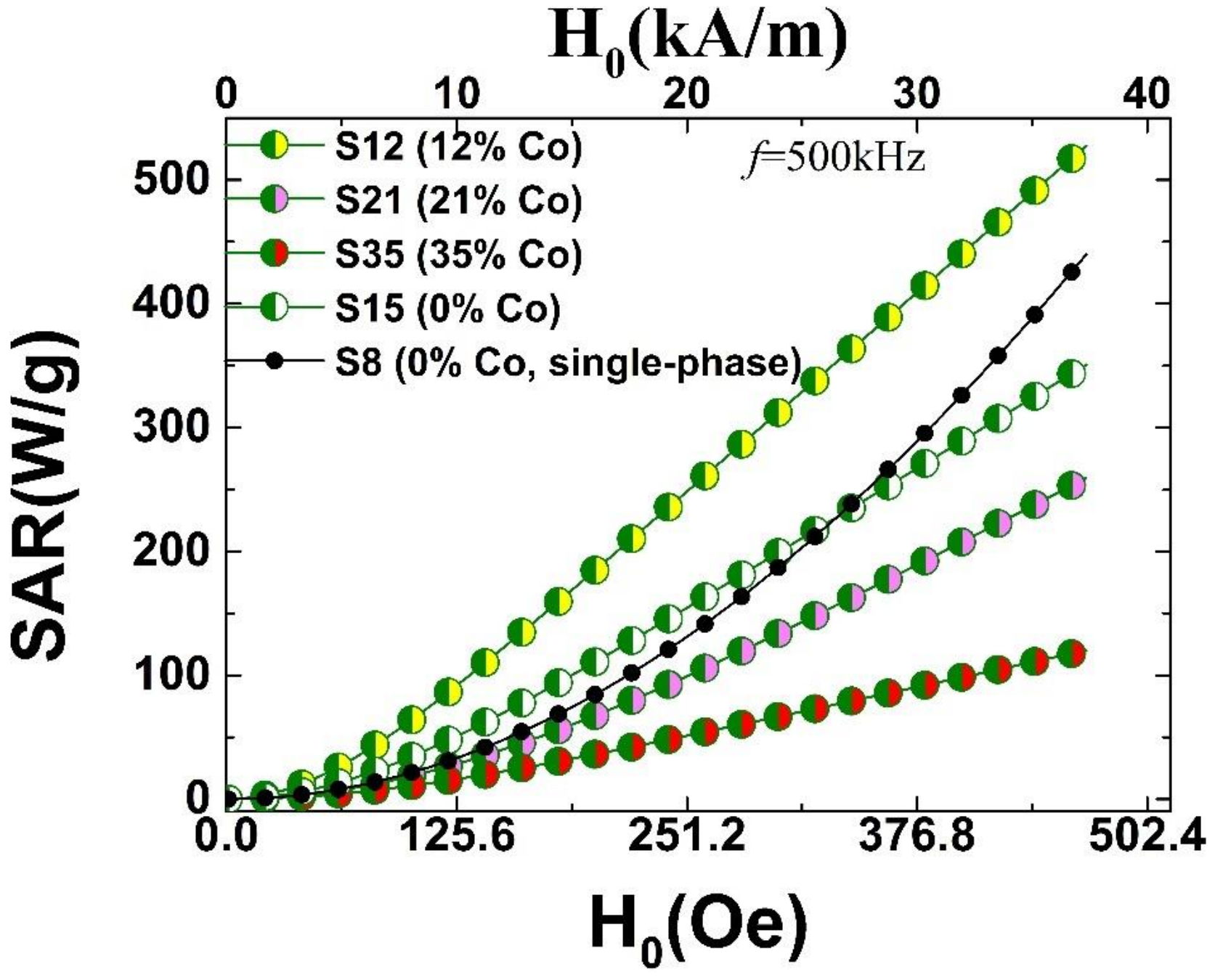

Fig. 12 Specific Absorption Rate (SAR) calculated on the grounds of susceptibility losses (AC field amplitude, $H_0$ and $f$ = 500 kHz) for Co-substituted heterostructured nanocrystals (15 nm), which are compared with 8 nm (S8) and 15 nm (S15) Fe-containing only[34] nanocrystals; SAR values are derived in view of to the linear response theory of a modified Néel-Brown relaxation Monte Carlo model.

# SUPPLEMENTARY INFORMATION

## Tailoring defects and nanocrystal transformation for optimal heating power in bimagnetic $Co_yFe_{1-y}O@Co_xFe_{3-x}O_4$ particles

George Antonaropoulos,[a,b] Marianna Vasilakaki,[c] Kalliopi N. Trohidou,[c] Vincenzo Iannotti,[d] Giovanni Ausanio,[d] Milinda Abeykoon [e], Emil S. Bozin [f] and Alexandros Lappas *[a]

a. *Institute of Electronic Structure and Laser, Foundation for Research and Technology - Hellas, Vassilika Vouton, 71110 Heraklion, Greece*

b. *Department of Chemistry, University of Crete, Voutes, 71003 Heraklion, Greece*

c. *Institute of Nanoscience and Nanotechnology, National Center for Scientific Research Demokritos, 15310 Athens, Greece*

d. *CNR-SPIN and Department of Physics "E. Pancini", University of Naples Federico II, Piazzale V. Tecchio 80, 80125 Naples, Italy*

e. *Photon Sciences Division, National Synchrotron Light Source II, Brookhaven National Laboratory, Upton, New York 11973, USA*

f. *Condensed Matter Physics and Materials Science Division, Brookhaven National Laboratory, Upton, New York 11973, USA*

*Corresponding author: lappas@iesl.forth.gr

## Table of Contents

**S1. Monte Carlo calculation of $H_{EB}$ and $H_C$**

In order to give a deeper insight on the effect of Co substitution in combination with the defects in Model #1, where the core/shell ratio is 25%/75% and the defects are located only in the core, we considered nanoparticles described by variants of Model #1: a) without defects (def=0%), b) without Co (substitution = 0%), c) without defects and Co (def=0% and substitution = 0%) and we compare these results with the starting model (Co=10% and def=40% in the core and core IF). Figure S1 presents the MC results of the $H_{EB}$ and the Hc as a function of the cooling field.

In the case of the model variant, where Co is absent (blue line), the bonds at the interface become more ordered than those of starting Model #1 (black line) and this enhances the competition between the soft and the hard component at the core/shell (AFM/FiM) interface, increasing the exchange bias field. In the model variant where Co exists, but no defects are present (red line), the soft component along the interface is marginally enhanced, resulting to a small increase of $H_{EB}$. Finally, when both substitution and defects are absent (green line), we observe an imperceptible decrease of the $H_{EB}$. In all cases the absence of pinning centers either in the core (Co and defects) or the shell (Co sites) makes the $H_C$ noticeably diminishing. Thus, it seems that the enhancement of the $H_{EB}$ and $H_C$ values can be achieved with a relatively small, defected AFM core in a combination with a low Co percentage. Taking into account the calculated values for a similarly-sized, non-substituted core@shell particle ("pure" in Fig. 11b, main text), it seems that given an ideal core@shell particle with no defects at all in the shell, any combination of defects and Co% in the core would result to a similar total magnetic response, if not higher. The only way to noticeably lower the $H_C$ value, is to create a perfectly ordered FeO core, with no defects and Co sites (green line). This is though impossible to be achieved experimentally, due to the highly sub-stoichiometric nature of $Fe_{1-x}O$. Therefore, we conclude that substituting a particle's core only, is not followed by severe changes in the total magnetic response, apart from the stabilization of the core (protection against oxidation), which is obviously a key element for the emergence of exchange interactions at the interface. Since chemical substitution in our synthetic protocol is leading to Co incorporation in the shell as well, a similar analysis to investigate the combined actions of Co-substitution and defects in the shell only, has been done.

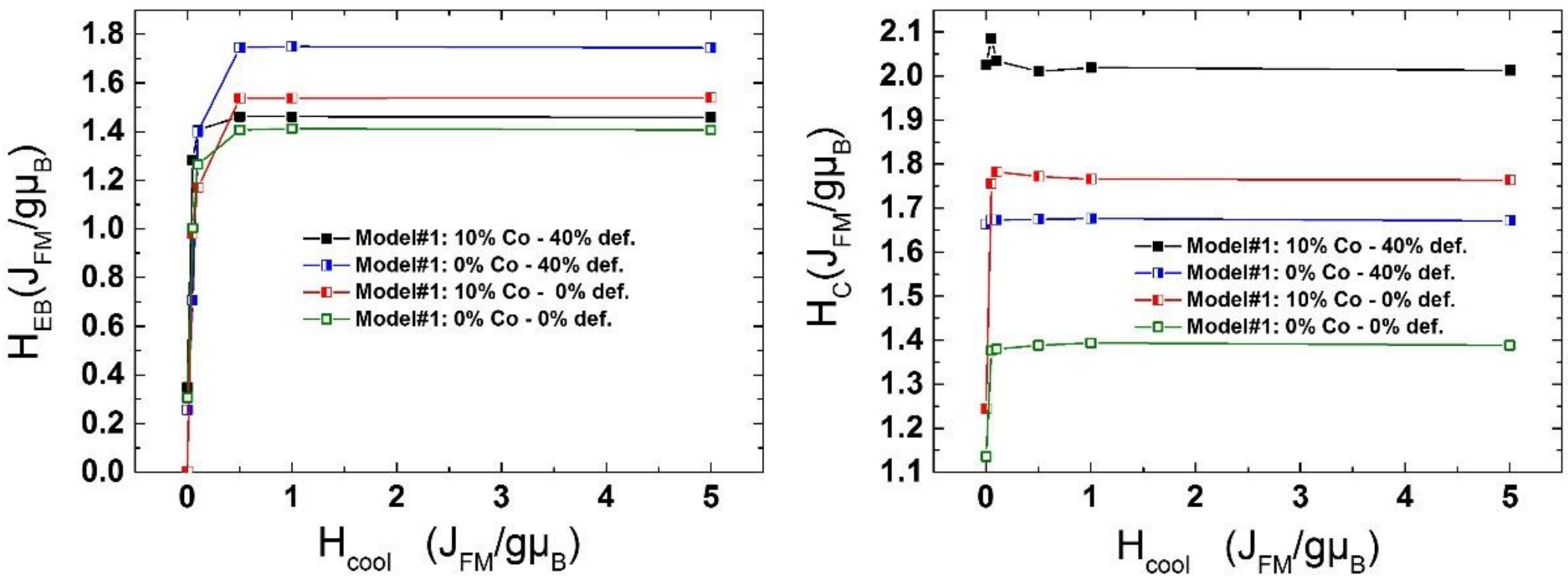

Figure S 1. MC simulation results for exchange bias, $H_{EB}$ and coercive field, $H_C$ for different simulated cooling fields, $H_{Cool}$ for variations in composition of model #1, as detailed in the graph legends. While keeping the morphological characteristics of the simulated nanoparticles the same, we tested different combinations of Co substitution and defects (vacant crystallographic sites). The lines connecting the points, extracted from the calculations, are a guide to the eye.

This was performed by simulating different variants of model #2, in which the core/shell ratio is 50%/50% and the defects are located in the shell only. MC simulated results for the cooling field dependence of the $H_{EB}$ and $H_C$ are shown in Figure S2 for the cases of model #2 a) without defects (red line), b) without Co (blue line), c) without Co and defects (green line) and d) an additional variant of reduced population of defects (orange line). The results are again compared with those for the starting model (Co=35% and def=40% in the shell) shown with the black line.

Interestingly, when comparing the 2 Co-substituted nanoparticle model variants with defects (black and yellow line) and the one without defects in its structure (red line), as the population of defects becomes smaller and smaller and finally drops to zero, the fraction of defects at the shell IF also diminishes, resulting in a stronger shell-IF magnetization component. Thus the competition with the core-IF is raising and the exchange bias gradually increases too. This does not affect the $H_C$ which remains almost the same, due to the contribution of the large fractions of Co sites in the magnetocrystalline anisotropy. When Co and defected sites are absent (green line) in an ideal perfectly ordered spinel ferrite, this seems to negatively affect the effective shell interface anisotropy and further weakens the competition between the two phases at the interface, decreasing the $H_{EB}$, but not considerably. A high rate of defects (40%), without any Co sites (blue line), worsens the situation even more, since the shell-IF magnetization component seems unable to maintain a proper magnetic ordering to oppose the AFM core. The depletion of the magnetocrystalline anisotropy when the Co sites are absent, also decreases the $H_C$ (blue and green line in Figure S2, right panel). Overall, the optimum magnetic behaviour in this case (core/shell = 50%/50%) can be achieved with a non-defective and highly Co substituted FiM shell. The $H_{EB}$ values are in the case of model #2 significantly lower than those in model #1 and the pure, non-substituted model due to the lower volume fraction of FiM shell, giving the AFM core the opportunity to prevail.

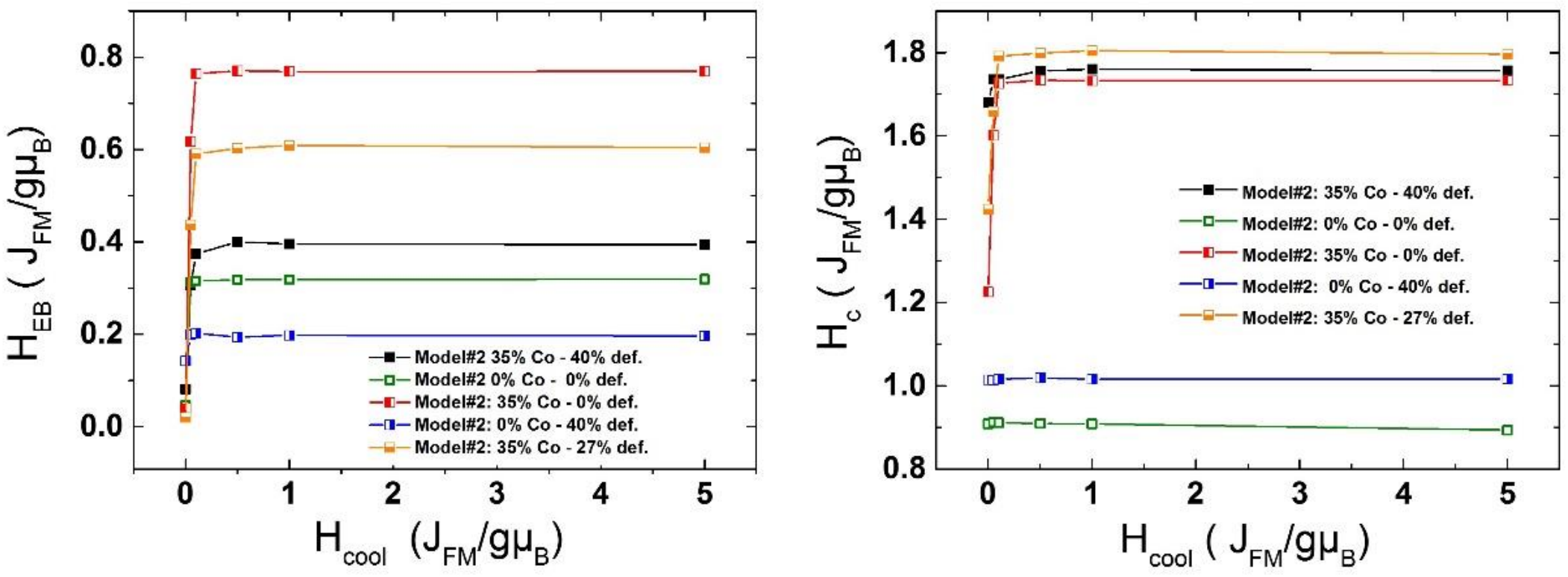

Figure S 2. MC simulation results for exchange bias, $H_{EB}$ and coercive field, $H_C$, for different simulated cooling fields, $H_{Cool}$ for variations of composition in model #2, as detailed in the graph legends. We keep again the morphological characteristics of the simulated nanoparticles the same and test different combinations of Co-levels and defects (vacant crystallographic sites). The lines connecting the points, extracted from the calculations, are a guide to the eye.

**S2. Calculation of the Specific Absorption Rate (SAR)**

The above calculations of Hc and $H_{EB}$ as a function of the cooling field were performed for low temperature (as in the experiments). For the following calculations, the Néel-Brown relaxation model [1,2] is used to calculate the SAR due to susceptibility losses. Therefore, for our calculations the experimental values at T=313 K have been used.

The Specific Absorption Rate is expressed as: $SAR(f) = \frac{\mu_0 \pi f \chi'' H_0^2}{\rho}$

where ρ: average density of each ferrite nanoparticle equals to Vcore/Vtot × $\rho_{Fe1-yCoyO}$ + Vshell/Vtot × $\rho_{Fe3-yCoyO4}$ and $\rho_{Fe1-yCoyO}$ = 8×$M_W$/0.602×$\alpha^3_{Fe1-yCoyO}$ and $\rho_{Fe3-yCoyO4}$ = 4×MW/0.602×$\alpha^3_{Fe3-yCoyO4}$ with $M_W$ : molecular weight and α : lattice constant

$H_0$ : AC field amplitude, and $f$ : field frequency,

$\chi''$: imaginary part of the complex susceptibility that involves the effective relaxation times for the two absorption mechanisms, namely, Brown ($\tau_B$) and Néel ($\tau_N$).

Temperature is set to T= 313 K.

η : the medium viscosity with a value of 0.65 x $10^{-3}$ Pa.s (approximately the value for water at 40°C)

ϕ : volumetric ratio of the NPs set to 0.001

A surfactant layer that covers the nanoparticles (NPs) is taken to have a thickness of 4 nm that is a parameter introduced in the calculation for the Brownian relaxation time.[3]

In our calculations the effective nanoparticle anisotropy constant is taken from the equation: $(KV)_{eff} = 25k_BT_B$ where $T_B$ is the blocking temperature extracted from the experimental ZFC magnetization curves and V the volume of the particles. The saturation magnetization $M_S$ at 300K is extracted from the experimental hysteresis loops of the Co-substituted nanoparticles. For comparison purposes, we considered also the experimental values for the defected spherical nanoparticles named S15 and the S8 magnetite NPs of D= 15 and 8 nm respectively.[3]

Exploring the effect of AC field strength, the SAR according to the Linear Response Theory for the Néel-Brown relaxation model,[1] was calculated at different field amplitudes, $H_0$, with a frequency of $f$= 500 kHz. The difference in the magnitude of SAR (due to susceptibility losses) of Co substituted and defected compared to pure and defected NPs is depicted in the plot shown in Fig. 12. The SAR ($H_0$) curves follow a trend similar to that in experimental findings previously reported.[4] The higher anisotropy volume of the core/shell nanoparticles results in the deviation from the quadratic field dependence of the SAR curve, which the smaller particle S8 only follows.[5]

Table S 1. Number and type of spins in each nanoparticle region, used for Monte Carlo calculations for the three heterostructured nanocrystal models, illustrated in Fig. 1 (main text). The corresponding relative volume fractions and Co percentages are also noted for quick reference.

| | | Total | Pure sites | Defects | Doped site | % Volume | % Doping |
|---|---|---|---|---|---|---|---|
| pure | Core | 257 | 196 | 61 | 0 | 20 | 0 |
| | Core-IF | 330 | 319 | 11 | 0 | | |
| | Shell-IF | 362 | 351 | 11 | 0 | 80 | |
| | Shell | 2170 | 1475 | 695 | 0 | | |
| model #1 | Core | 461 | 170 | 284 | 7 | 25 | 10 |
| | Core-IF | 338 | 224 | 36 | 78 | | |
| | Shell-IF | 558 | 498 | 0 | 60 | 75 | |
| | Shell | 1762 | 1585 | 0 | 177 | | |
| model #2 | Core | 1021 | 658 | 0 | 363 | 50 | 35 |
| | Core-IF | 530 | 349 | 0 | 181 | | |
| | Shell-IF | 654 | 261 | 55 | 338 | 50 | |
| | Shell | 914 | 124 | 577 | 213 | | |

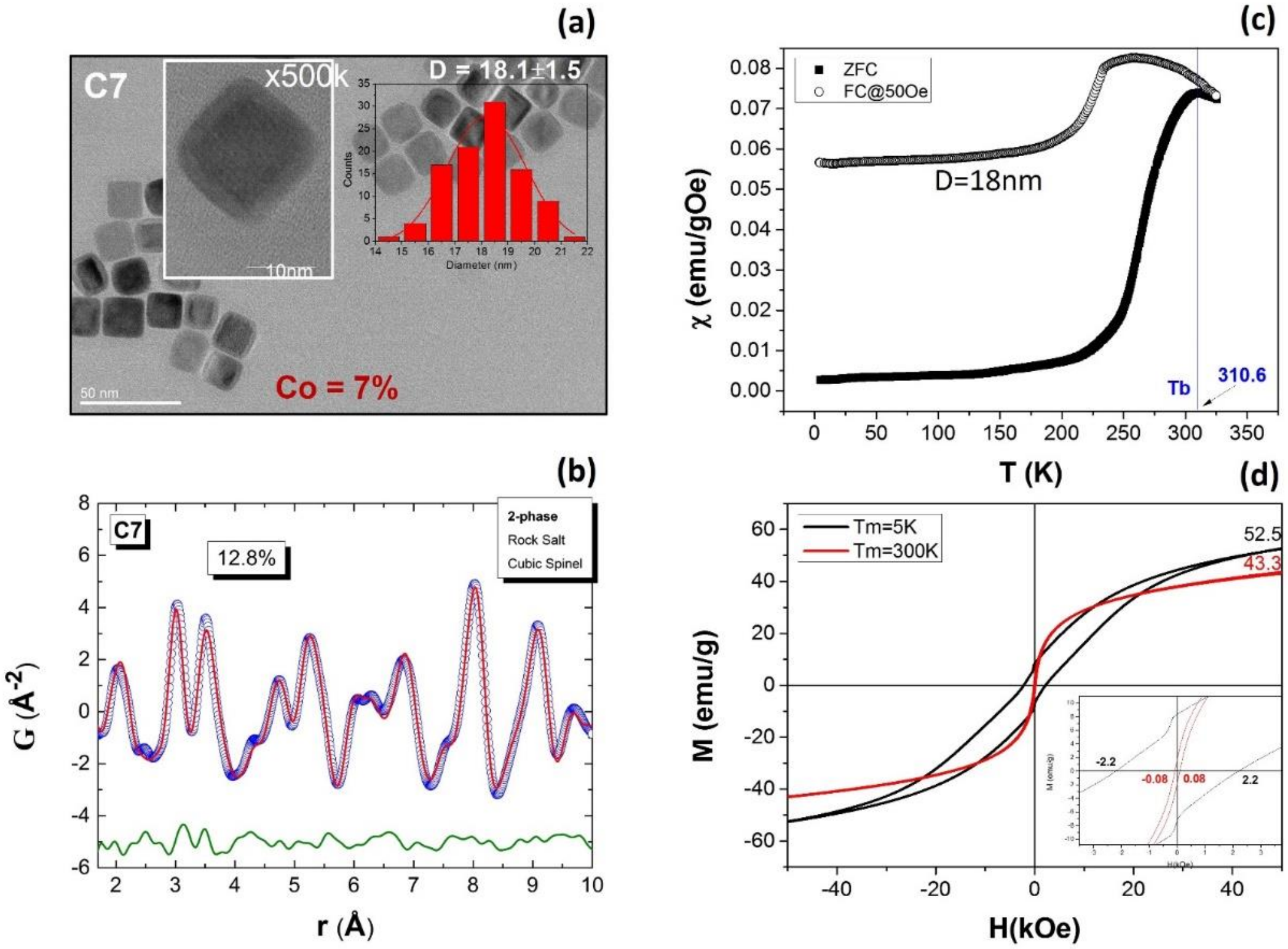

Figure S 3. Cubic Sample C7 (7% Co): (a) Low-magnification bright field TEM images of a cubic sample, entailing an edge length of ~18 nm and inset of HR-TEM showing a clear core@shell structure. (b) xPDF fit of data at T= 300 K over the low-r PDF region (1 nm) for the same sample, assuming a 2-phase rock-salt/cubic spinel model (Fm-3m/ Fd-3m, $R_w$= 12.8%). (c) Temperature evolution of the zero-field cooled (ZFC, solid line) and field-cooled (FC, dotted line) susceptibility curves), under a magnetic field of 50 Oe. (d) The low-field part of the hysteresis loop at 5 K and 300 K, taken under zero- and field-cooled ($H_{cool}$ = 50 kOe) protocols.

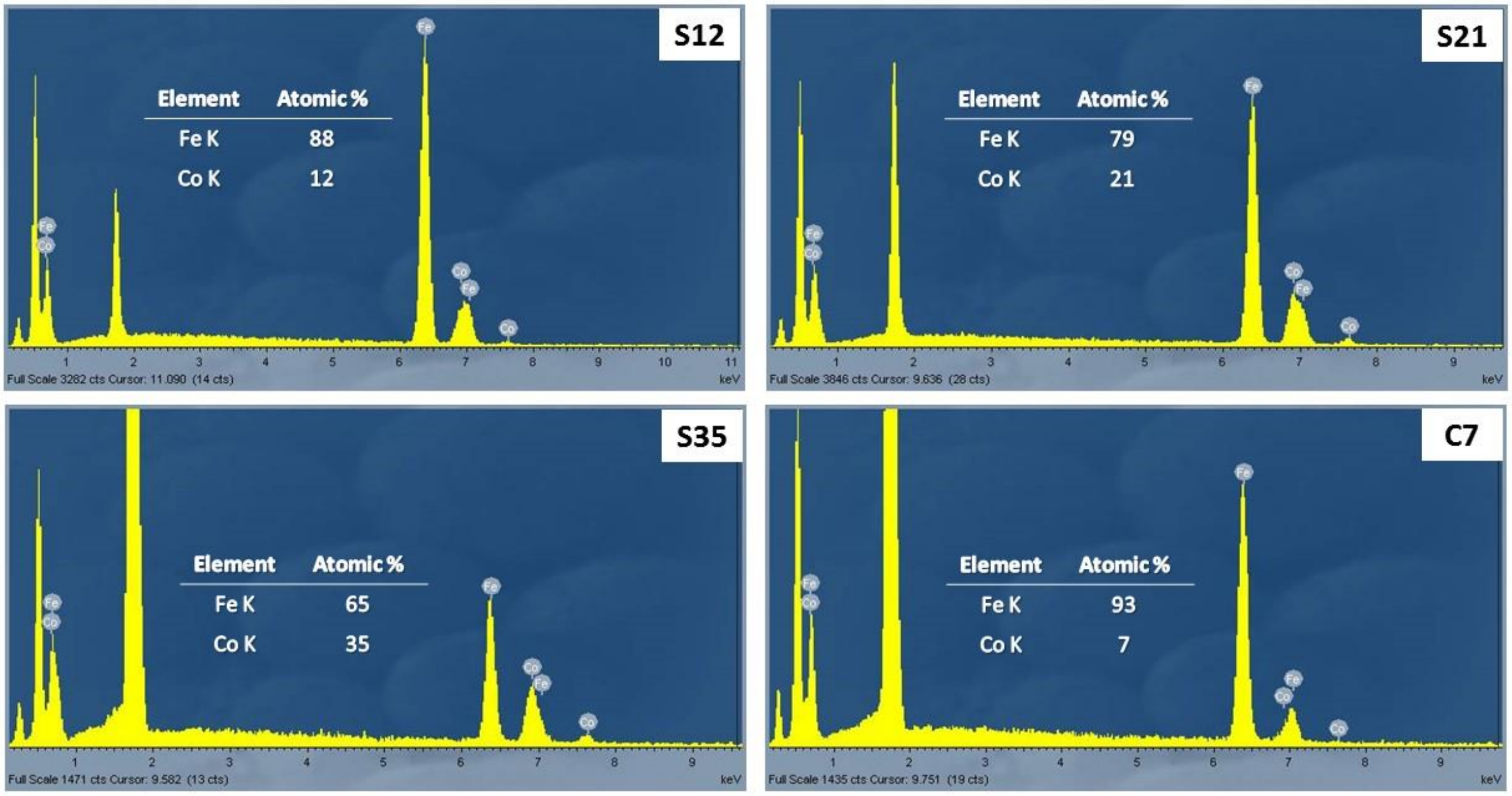

Figure S 4. SEM-EDS spectra showing the characteristic K-peaks of Fe and Co of the three studied spherical nanocrystal samples (S12, S21, S35) and one cubic (C7). Their relative atomic % abundance is automatically calculated by INCA software. The peak at ~1.8 keV is coming from the Si substrate and this is why it is in some case too high compared to Fe and Co peaks.

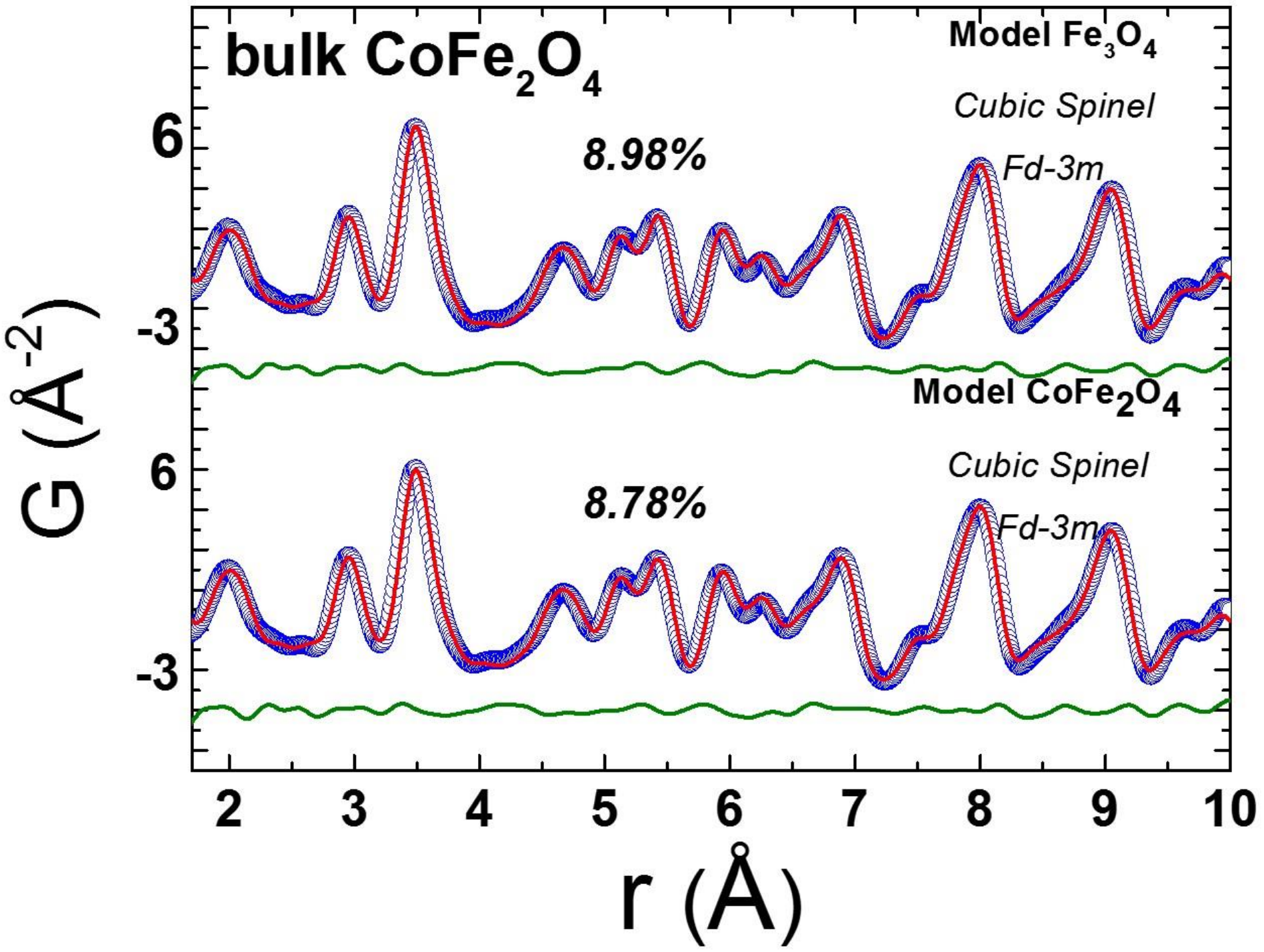

Figure S 5. xPDF fits of data at T= 300 K over the low-r PDF region (1 nm) for bulk reference $CoFe_2O_4$ assuming a $Fe_3O_4$, cubic spinel model (Fd-3m, $R_w$= 8.98%) at the top and a $CoFe_2O_4$ cubic spinel model (Fd-3m, $R_w$= 8.78%) at the bottom, where half of the Fe-ions in the octahedral sites of the model shown at the top, have been replaced by Co, resembling the simplest cobalt ferrite cubic spinel structure. These two models are practically equally good in describing the structure, since xPDF cannot distinguish between Fe and Co ions.

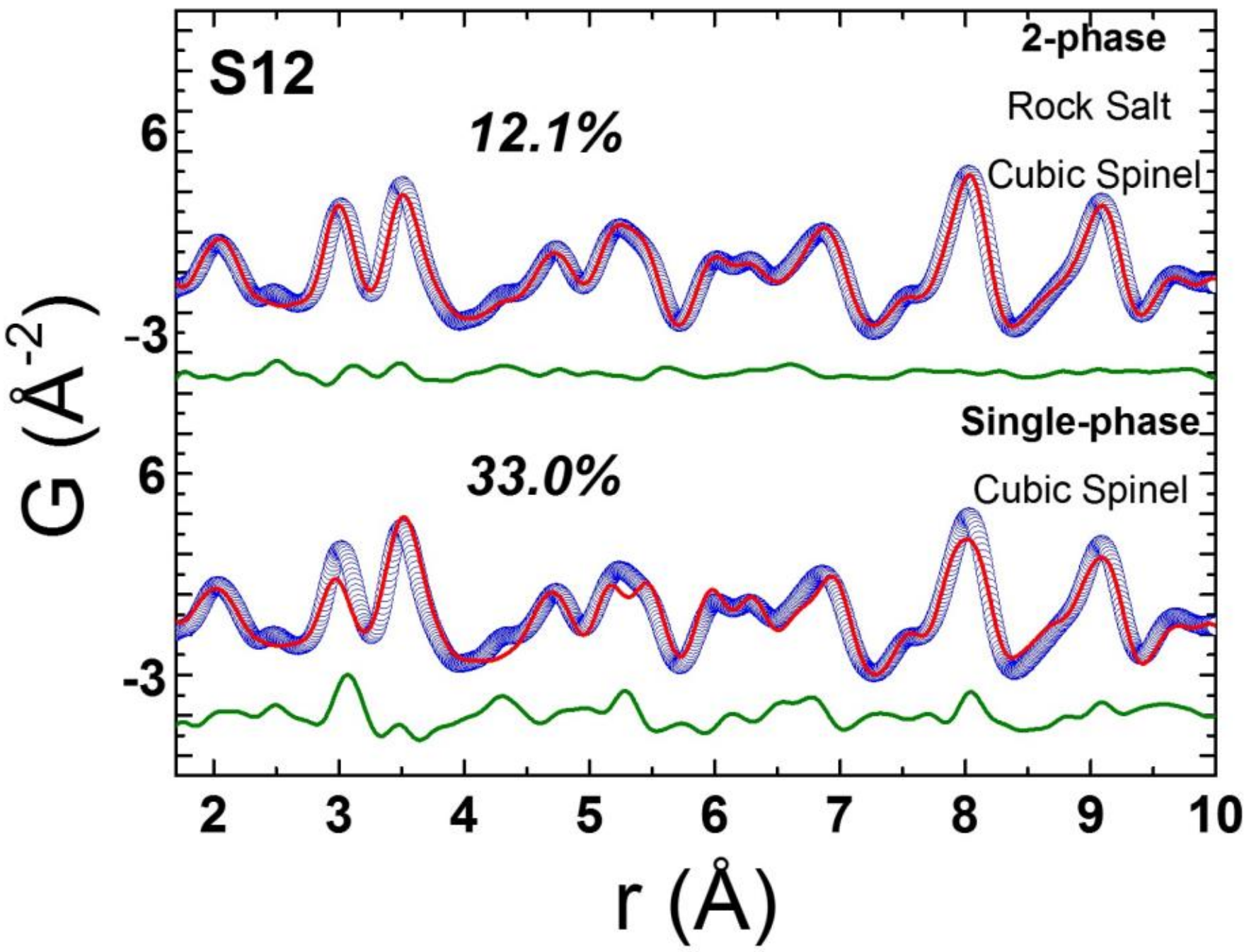

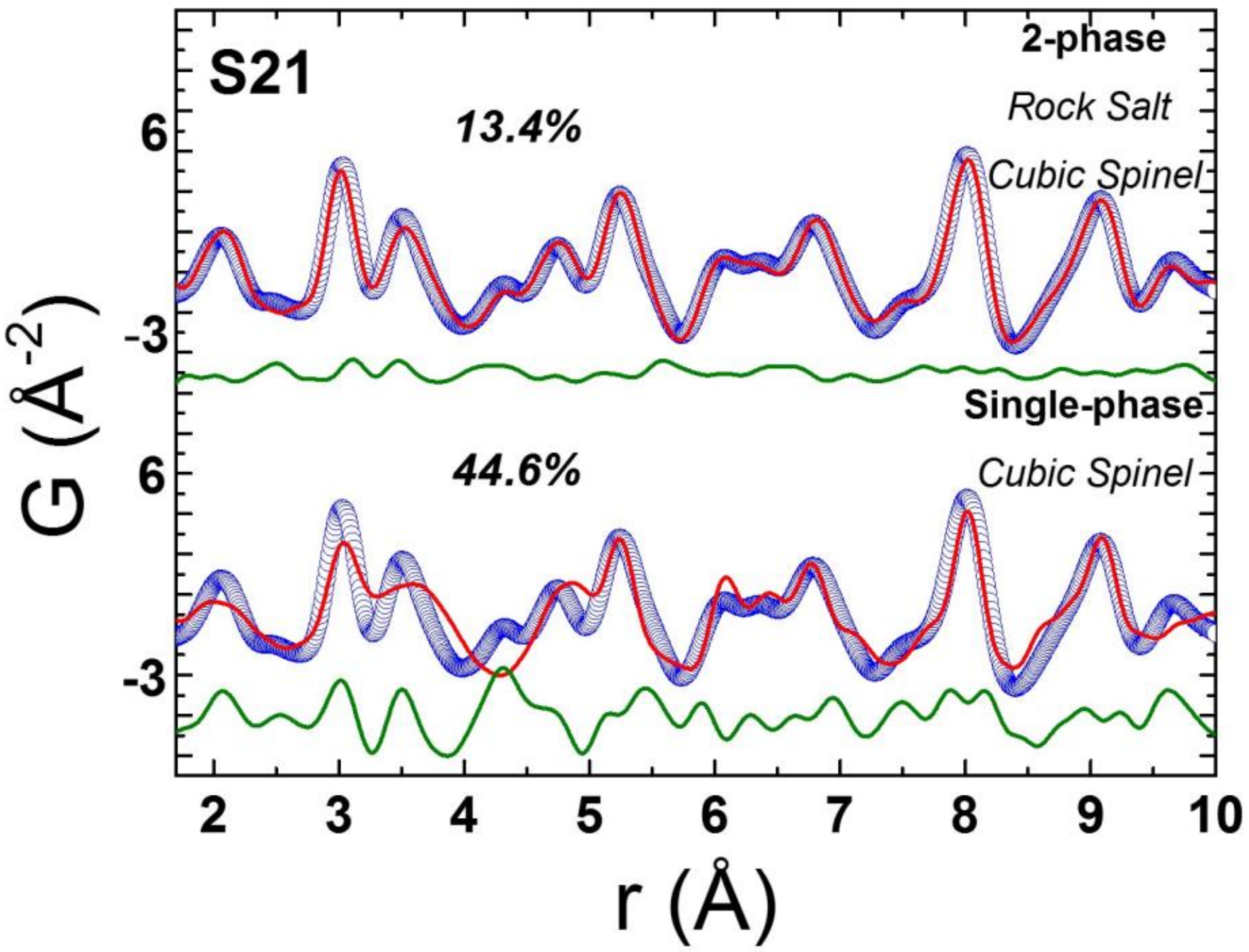

Figure S 6. xPDF fits of data at T= 300 K over the low-r PDF region (1 nm) for sample S12 (top, $R_w$= 33.0% *vs.* 12.1%) and S21 (bottom, $R_w$= 44.6% *vs.* 13.4%), assuming either a single-phase, spinel-only model or a 2-phase rock-salt/cubic spinel model (Fm-3m/ Fd-3m).

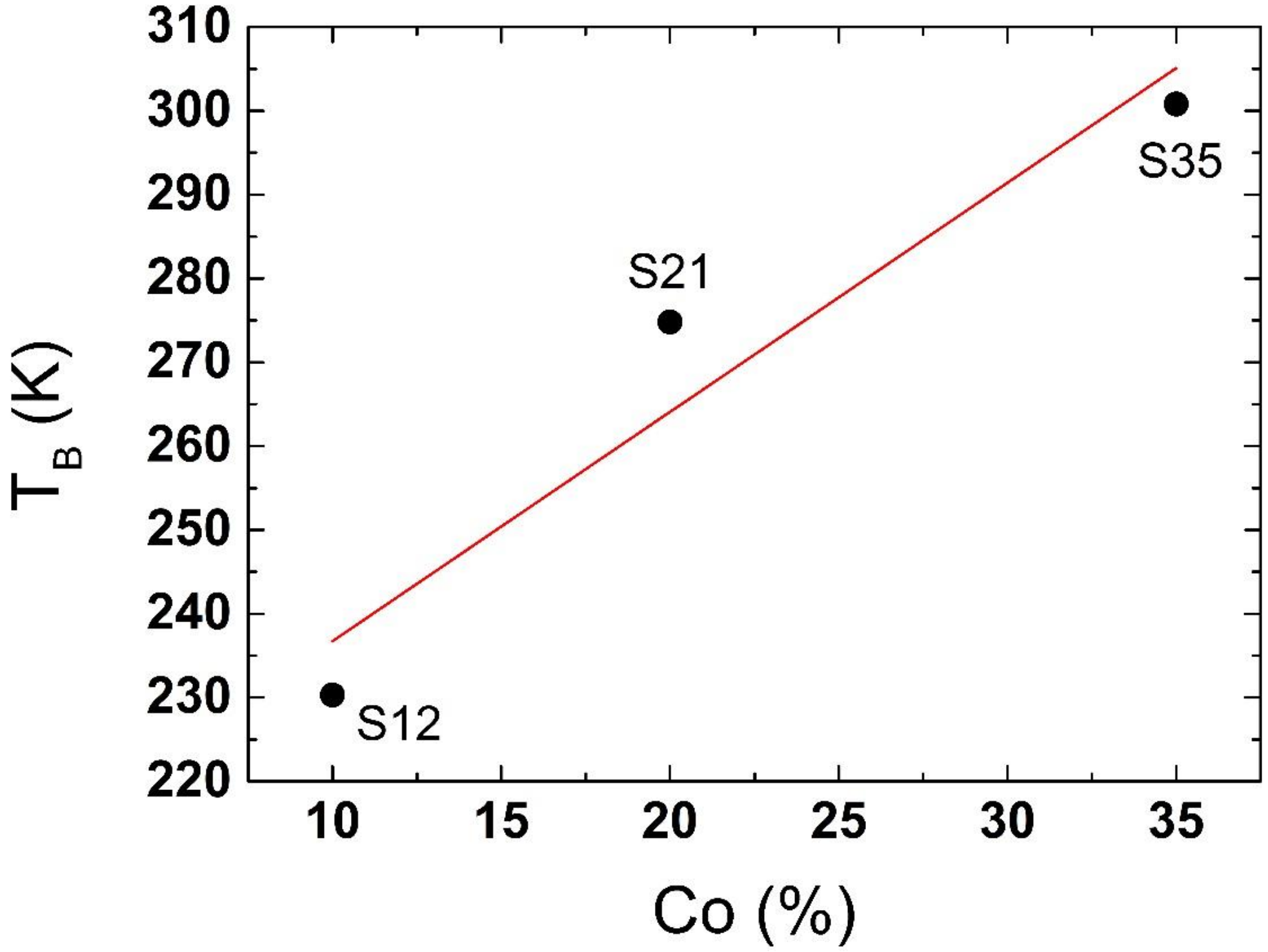

Figure S 7. Dependence of the blocking temperature ($T_B$) as extracted from the maximum of the zero-field cooled (ZFC) magnetic susceptibility curves (Fig. 9 – main text) for the spherical nanocrystal samples S12, S21 and S35, showing a nearly linear increase of $T_B$, as Co-content increases.

Table S 2. Crystallographic parameters of the cubic spinel and rock salt models utilized in the refinements of the low-r atomic PDF (r=1-10 Å) and parameters for bulk reference samples derived from fitting the low-r region of their atomic PDF.

| | Models | | refined parameters: Reference Samples | |
|---|---|---|---|---|
| | Spinel[*] | Rock Salt[**] | Bulk $CoFe_2O_4$ | Bulk FeO |
| Symmetry | Fd-3m | Fm-3m | Fd-3m | Fm-3m |
| a (Å) | | | | |
| b (Å) | 8.397 | 4.3108 | 8.37699(4) | 4.32151(5) |
| c (Å) | | | | |
| V ($Å^3$) | 592.0692 | 80.1076 | 587.8466(1) | 80.7061(1) |
| **Fe-Td** | | Uiso ($Å^2$) 0.003 | | Uiso ($Å^2$) 0.0120(5) |
| 8α | | | | |
| x=y=z | 0.125 | | 0.125 | |
| Uiso ($Å^2$) | 0.003 | | 0.0068(1) | |
| **Fe-Oh** | | x = y = z 0 | | x = y = z 0 |
| 16d | | | | |
| x=y=z | 0.5 | | | |
| Uiso ($Å^2$) | 0.003 | | 0.0074(1) | |
| **O** | | | | |
| 32e | | | | |
| x=y=z | 0.2551 | 0.5 | 0.5 | 0.5 |
| Uiso ($Å^2$) | 0.003 | 0.003 | 0.0074(1) | 0.0229(6) |
| % vol. fraction | ... | ... | 100 | 100 |
| Rw (%) | ... | ... | 9.0 | 14.4 |

* see ref[6]

** see ref[7]

Table S 3. Parameters for the nanocrystal samples, derived from fitting the low-r region of their atomic PDF (r= 1-10 Å) and $\sigma_0$ values extracted from the Einstein fit of their T-dependent isotropic temperature factors, giving the static disorder of the system.

| | refined parameters | | | | | |
|---|---|---|---|---|---|---|
| | S12 | | S21 | | S35 | |
| | Spinel | Rock salt | Spinel | Rock salt | Spinel | Rock salt |
| Symmetry | Fd-3m | Fm-3m | Fd-3m | Fm-3m | Fd-3m | Fm-3m |
| a (Å) | | | | | | |
| b (Å) | 8.415(1) | 4.305(1) | 8.407(5) | 4.296(2) | 8.411(6) | 4.295(1) |
| c (Å) | | | | | | |
| V (Å³) | 595.885(1) | 79.785(1) | 594.187(1) | 79.285(1) | 595.035(1) | 79.230(1) |
| **Fe-Td** | | | | | | |
| 8α | | | | | | |
| x=y=z | 0.125 | | 0.125 | | 0.125 | |
| Uiso (Å²) | 0.0088(1) | Uiso (Å²) 0.0096(1) | 0.0151(6) | Uiso (Å²) 0.0098(1) | 0.019(1) | Uiso (Å²) 0.0087(1) |
| **Fe-Oh** | | | | | | |
| 16d | | | | | | |
| x=y=z | 0.5 | x = y = z 0 | 0.5 | x = y = z 0 | 0.5 | x = y = z 0 |
| Uiso (Å²) | 0.0086(1) | | 0.0118(4) | | 0.0101(5) | |
| **O** | | | | | | |
| 32e | | | | | | |
| x=y=z | 0.2551 | x=y=z 0.5 | 0.2551 | x=y=z 0.5 | 0.2551 | x=y=z 0.5 |
| Uiso (Å²) | 0.018(1) | Uiso (Å²) 0.0288(3) | 0.017(1) | Uiso (Å²) 0.0254(2) | 0.021(2) | Uiso (Å²) 0.0231(2) |
| % vol. fraction | 73.6(1) | 26.4(1) | 63.6(1) | 36.4(1) | 47.1(1) | 52.9(1) |
| Rw (%) | 12.1 | | 13.4 | | 10.2 | |
| Oh $\sigma_0$ | 0.00343(2) | ... | 0.00419(1) | ... | 0.00312(2) | ... |
| Td $\sigma_0$ | 0.00387(3) | ... | 0.00832(3) | ... | 0.00956(3) | ... |
| RS $\sigma_0$ | ... | 0.00377(2) | ... | 0.00324(2) | ... | 0.00242(2) |

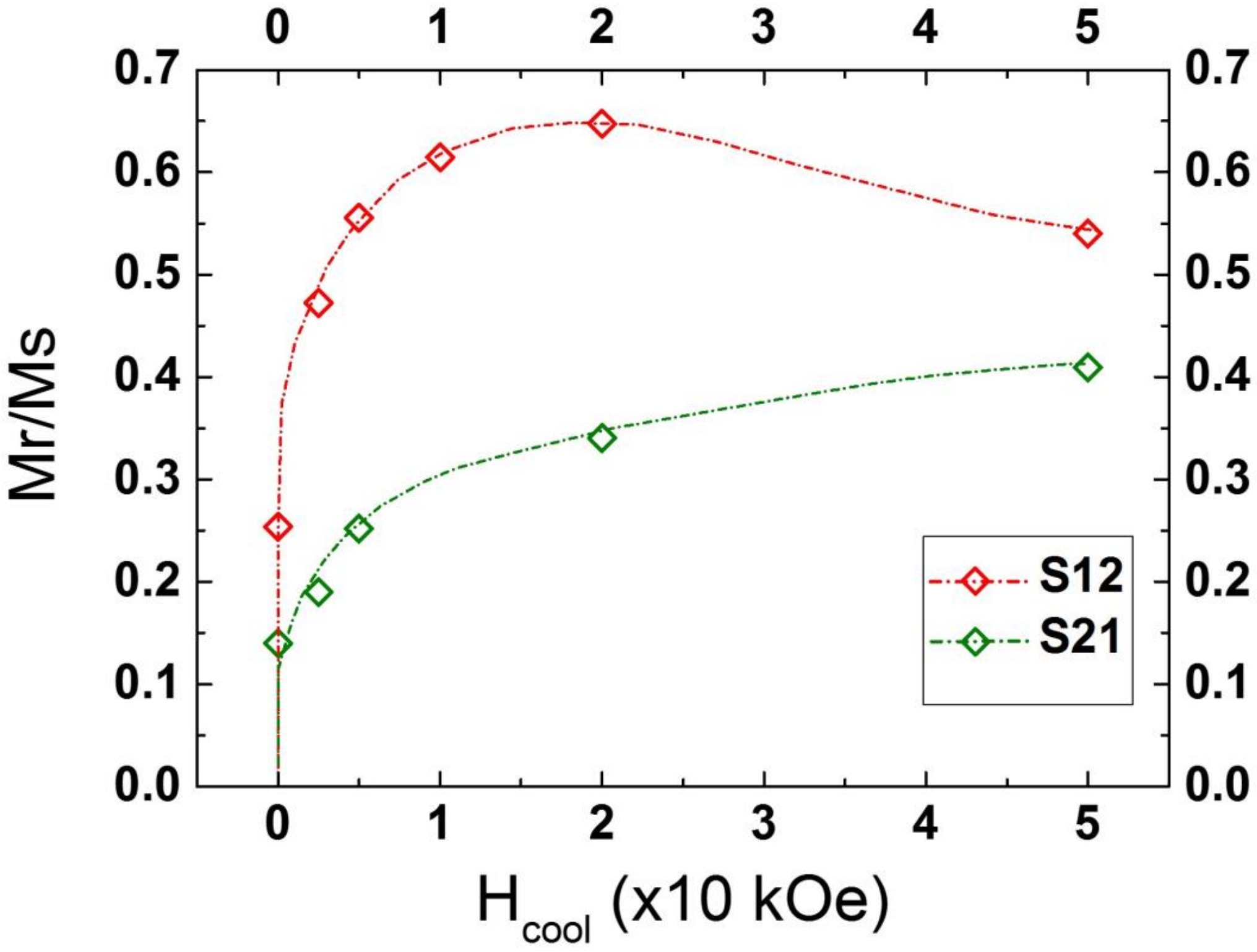

Figure S 8. The experimentally determined $M_r/M_S$ ratio obtained at varying cooling-field strengths ($H_{cool}$) for nanocrystal samples S12 (red symbols) and S21 (green symbols). The lines are a guide to the eye.

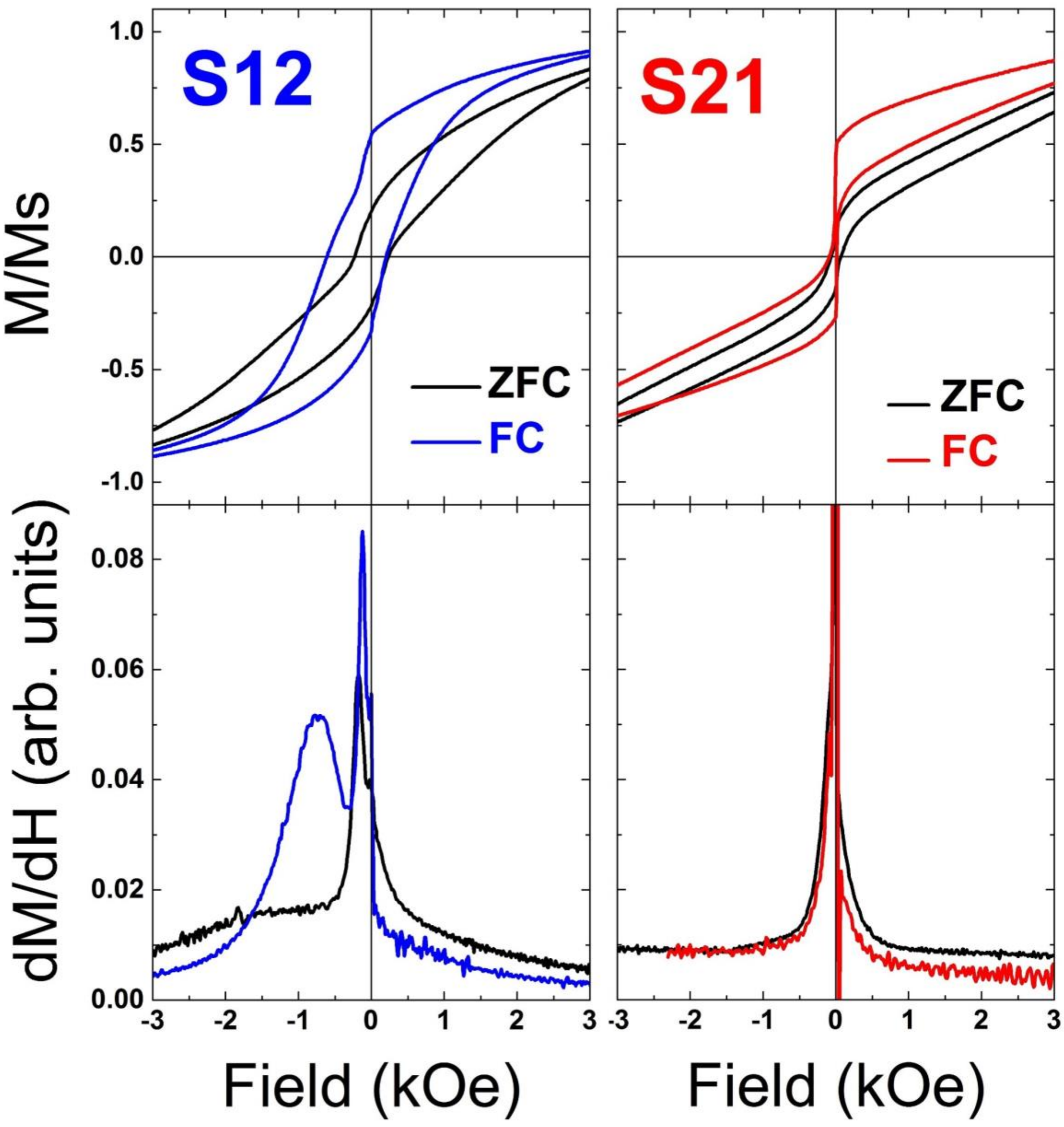

Figure S 9. The low-field, normalized hysteresis loop regions at 5 K, for samples S12 and S21, comparing the $M/M_S$ data measured after 50 kOe field-cooled (FC, colored curves) and zero-field cooled (ZFC, black curves) protocols. The panels beneath the normalized magnetization $M/M_S$ present the corresponding differential change (dM/dH) when switching from positive to negative field saturation.

Table S 4. Size and composition of nanocrystal samples, their blocking temperature $T_B$, as well as the saturation magnetization, $M_S$ and the coercive field, $H_C$ at T= 5 and 300 K. The field-cooled (FC) hysteresis loop (M-H; $H_{cool}$ = 50 kOe) characteristics, including exchange bias $H_{EB}$ at 5 K for samples S12 and S21.

| Sample | Size (nm) | Co% | $T_B$ (K) | $M_S$ (emu/g), 300K | $H_C$ (Oe), 300K | $M_S$ (emu/g), 5K | $H_C$ (Oe), 5K | $M_S$ (emu/g), FC curve, 5K | $H_C$ (Oe), FC curve, 5K | $H_{EB}$ (Oe) |
|---|---|---|---|---|---|---|---|---|---|---|
| S12 | 15.2 | 12 | 230 | 42.5 | 180 | 39.2 | 2350 | 50.7 | 2617 | 2022 |
| S21 | 13.9 | 21 | 275 | 22.2 | 265 | 17 | 702 | 19.7 | 823 | 367 |
| S35 | 17.8 | 35 | 301 | 14.8 | 335 | 12.4 | 1182 | / | / | / |
| C7 | 18.1 | 7 | 311 | 43.3 | 87 | 52.5 | 2188 | / | / | / |